\documentclass[a4paper]{paper_cw}

\usepackage{helvet}
\usepackage{amsmath}
\usepackage{amssymb}
\usepackage{mathrsfs}
\usepackage{graphics}
\usepackage{natbib}
\usepackage{color}
\usepackage{fancyhdr} 
\usepackage{colortbl} 
\usepackage{subfigure}
\usepackage[dvips]{graphicx} 
\usepackage{marvosym}
\usepackage{natbib}
\usepackage{import}

\long\def\symbolfootnote[#1]#2{\begingroup \def\thefootnote{\fnsymbol{footnote}}\footnote[#1]{#2} \endgroup}

\usepackage{graphicx}

\begin{document}

\title{An electromechanically coupled beam model for dielectric elastomer actuators}

\author{Dengpeng Huang \and Sigrid Leyendecker}

\institute{D. Huang(\Letter), S. Leyendecker  \at Institute of Applied Dynamics, Friedrich-Alexander-Universität Erlangen-Nürnberg, Germany
\\\email{dengpeng.huang@fau.de}}

\maketitle
\thispagestyle{empty}

\abstract{In this work, the Cosserat formulation of geometrically exact beam dynamics is extended by adding the electric potential as {an} additional degree of freedom to account for the electromechanical coupling in the Dielectric Elastomer Actuators (DEAs). {To be able to generate complex beam deformations via dielectric actuator}, a linear distribution of electric potential on the beam cross section is proposed. Based on this electric potential, the electric field and the strain-like electrical variable are {defined for the} beam, where the strain-like electrical variable is work-conjugated {to} the electric displacement. The electromechanically coupled strain energy for the beam is derived consistently from continuum electromechanics, which leads to the direct application of the material models in the continuum to the beam model. The electromechanically coupled problem in beam dynamics is first spatially semidiscretized by 1D finite elements and then solved via variational time integration. By applying different electrical boundary conditions, different deformations of the beam are obtained in the numerical examples, including contraction, shear, bending and torsion. The damping effect induced by the viscosity as well as the total energy of the beam are evaluated. The deformations of the electromechanically coupled beam model are compared with the results of the 3D finite element model, where a good agreement of the deformations in the beam model and that in the 3D finite element model is observed. However, less degrees of freedom are required to resolve the complex deformations in the beam model.}


\section{Introduction}
With the wide application of robotics in industrial production, medical treatment and daily life, better performances of robotic systems are demanded, such as {a} higher efficiency in energy, completing complex tasks and {a} safe interaction with environment. To cope with these challenges, the Dielectric Elastomer Actuators (DEAs) have been developed to serve as artificial muscles for soft robotics, see e.g. \citep{bar2000electroactive}, \citep{lowe2005dielectric}, \citep{kovacs2009stacked} and \citep{duduta2019realizing}. The DEA is essentially composed by multiple stacked capacitors where the dielectric elastomer is sandwiched between two compliant electrodes. When {an} external electric field is applied to the DEA, the dielectric material will be polarized{,} resulting in electrostatic pressure, see the models in \citep{pelrine1998electrostriction} , \citep{wissler2007electromechanical}, \citep{suo2008nonlinear} and \citep{schlogl2017polarisation} for instance. Due to the contractive pressure, the contraction of the DEA will be induced such that it can be applied as {an} actuator. The deformation behavior of the DEA is governed by the electromechanical coupling in the dielectric material.

For the general investigation of the electromechanical coupling behavior, much effort has been made to address the nonlinear electroelasticity in the past years, see e.g. the theory of interaction of electromagnetic and elastic fields in deformable continua in \citep{1978meto}, the nonlinear electroelasticity formulation for the finite deformation in \citep{dorfmann2005nonlinear} and the variational formulations of the electro- and magneto-elastostatics in \citep{vu2007numerical}. Additionally, material models of the dielectric elastomers have been investigated, see e.g. \citep{zhao2007electromechanical}, \citep{vu2007numerical} and \citep{suo2010theory}. In \citep{khan2013variational}, {a} viscoelastic effect is introduced to account for the damped dynamic behavior in silicon based dielectric {elastomers}. {A} viscoelastic 3D finite element model of the DEA is developed by \citep{schlogl2016electrostatic} for the dynamic analysis {using} a structure preserving {time} integration scheme. This model is extended to {flexible multibody system dynamics} in \citep{schlogl2016dynamic}.

The finite element models introduced above provide a powerful and accurate tool for solving the electromechanical coupling problem in DEA. However, huge {amounts} of degrees of freedom are required in large 3D finite element models, which leads to inefficient computation and difficulties for the optimal control of DEA. {Especially for long thin artificial muscles, 3D finite element model is more expensive in computational cost than beam model.} Further more, the coupling between 3D finite element {models} and rigid body {is possible but has to be specially addressed in multibody system}. The geometrically exact beam performs well concerning the tradeoff between computational cost and accuracy for the simulation of slender structures like the stacked DEA. By assigning the rotational degrees of freedom to points in continuum, the Cosserat formulation \citep{cosserat1909theorie} of geometrically exact beam closes the gap between classical continuum mechanics and rigid (multi-)body dynamics, which leads to a consistent {formulation} of flexible multibody systems. The fundamental formulations on geometrically exact beam can be found in \citep{simo1985finite} and \citep{antmannonlinear} for instance. The time integration of constrained geometrically nonlinear beam dynamics has been discussed {by many authors, see e.g. the energy conserving/decaying algorithms in \citep{ARMERO20012603}}, the energy-momentum scheme with null space method in \citep{betsch2006discrete} and the variational integrators in \citep{leyendecker2008variational}. To account for the electric field in {a} beam, the focus has been put on the piezoelectric effect, see \citep{krommer2002electromechanically}, \citep{tadmor2003electromechanical} and \citep{schoeftner2012electromechanically}. An analytical model for dielectric elastomer based microbeam is discussed in \citep{feng2011dynamic}.  However, the electromechanically coupled problem of dielectric elastomers in geometrically exact beam is still not given. 

The objective of this work is to develop an electromechanically coupled beam model for the simulation of stacked dielectric elastomer actuators, where the Cosserat formulation of geometrically exact beam dynamics is extended by adding the electric potential as the additional degree of freedom. A linear distribution of electric potential on the beam cross section is proposed to generate different beam deformations including contraction, shear, bending and torsion. The electric field in the beam is computed from the gradient of the electric potential. Based on the formulation of deformation gradient and electric field in the beam, the electromechanically coupled strain energy function for the beam is derived consistently from the strain energy function in {continuum electromechanics}, which leads to the direct application of the material models in continuum electromechanics to the beam model. The viscoelastic effect is taken into account in the non-conservative force term. The electromechanically coupled problem in beam dynamics is first semidiscretized with 1D spatial finite elements and then solved via variational time integration. By applying different electrical boundary conditions, different deformations of the beam are obtained in the numerical examples, which are compared with the results of the 3D finite element model.

This paper is structured as follows: In Section 2 and 3, the governing equations for electromechanical coupling in continuum electromechanics and geometrically exact beam are presented, respectively. Then the formulation of kinematic variables including the deformation gradient, the electric potential and the electric field are derived for the beam in Section 4. Section 5 presents the consistent derivation of {a} strain energy function for the beam from continuum electromechanics. In Section 6, the electromechanical coupling problem is solved within the variational time integration scheme with null space projection. The numerical examples of the developed model are presented in Section 7, followed by the conclusions in Section 8.

\section{Governing equations for electromechanical coupling in continuum electromechanics}
The finite deformation of a dielectric elastic solid occupying the domain $B \subset \mathbb{R}^3$ is distinguished by the initial and current configurations. The boundary of the solid $\partial B$ is composed by the Dirichlet type sections $\partial_{u} B$ and $\partial_{\phi} B$, and the Neumann type sections $\partial_{\sigma} B$ and $\partial_D B$.  By denoting the position of a material point in the initial configuration with $\mathbf{X}\in B_0$, the position of the material point in the current configuration at time $t$ is given by
\begin{align}
\mathbf{x}=\mathbf{X}+\mathbf{u}(\mathbf{X},t),
\end{align}
where $\mathbf{u}$ is the displacement. The deformations within the body induced by the electric field satisfy the balance law of momentum and the Maxwell equations.
\subsection{Balance of linear and angular momentum}
The local balance law of linear momentum in the dynamic process is given by
\begin{align}
\nabla_{\mathbf{X}} \cdot \mathbf{P} + \rho_0 \bar{\mathbf{b}} =\rho_0\mathbf{\ddot{u}} \qquad {\rm in} \quad B, \label{blm}
\end{align}
{subject to the Dirichlet and Neumann boundary conditions}
\begin{align}
\mathbf{u}&=\bar{\mathbf{u}} \qquad {\rm on} \quad \partial_{u} B,\\
\mathbf{P}\cdot \mathbf{N}&=\bar{\mathbf{T}} \qquad {\rm on} \quad \partial_{\sigma} B, \label{PNT}
\end{align}
where $ \mathbf{P} $ is the first Piola-Kirchhoff stress tensor, $\rho_0$ is the mass density in initial configuration, $\bar{\mathbf{b}}$ is the body force vector, $\mathbf{\ddot{u}}$ is the acceleration, {$\bar{\mathbf{u}}$ is the prescribed displacement and $\bar{\mathbf{T}}$ is the prescribed traction}. The local balance of angular momentum reads
\begin{align}
\mathbf{F}\mathbf{P}^T = \mathbf{P}\mathbf{F}^T,
\end{align}
in which $\mathbf{F}$ is the deformation gradient defined as $\mathbf{F}=\partial\mathbf{x}(\mathbf{X},t)/ \partial \mathbf{X}$.
\subsection{Maxwell equations}
By neglecting the magnetic field, the Maxwell equations are given by
\begin{align}
\nabla_{\mathbf{X}} \times \mathbf{E}^e=\mathbf{0}, \;\;\;\; \nabla_{\mathbf{X}} \cdot \mathbf{D}=0 \qquad {\rm in} \quad B \label{ED},
\end{align}
{subject to the Dirichlet and Neumann boundary conditions}
\begin{align}
\phi&=\bar{\phi} \qquad {\rm on} \quad \partial_{\phi} B,\\
\mathbf{D}\cdot \mathbf{N}&=\bar{Q} \qquad {\rm on} \quad \partial_D B, \label{DNQ}
\end{align}
with $ \mathbf{E}^e$ the electric field, $\mathbf{D}$ the electric displacement in the initial configuration, { $\phi$ the electric potential, $\bar{\phi}$ the prescribed electric potential, $\mathbf{N}$ the outward unit normal vector and $\bar{Q}$ the prescribed charges per unit area on the boundary $\partial_D B$}. The Eq. (\ref{ED})$_1$ leads to the definition of the electric field as the gradient of a scalar electric potential
\begin{align} 
\mathbf{E}^e=-\frac{\partial \phi}{\partial \mathbf{X}}. \label{E}
\end{align}
\subsection{Electromechanical coupling}
When the external electric field is imposed in the body of dielectric elastomer, the contractive pressure will be induced due to the polarization effects and thus the {deformation} of the body will be generated. The coupling effect between the electric field and the mechanical deformation is described by the strain energy function $\Omega(\mathbf{F}, \mathbf{E}^e)$ of the dielectric material in the constitutive equations 
\begin{align}
\mathbf{D}=-\rho_0\frac{\partial \Omega(\mathbf{F}, \mathbf{E}^e)}{\partial \mathbf{E}^e},  \;\;\;\;  \mathbf{P}=\rho_0\frac{\partial \Omega(\mathbf{F}, \mathbf{E}^e)}{\partial \mathbf{F}}. \label{CE}
\end{align}
For the dielectric materials, the electromechanical coupling can be described by the strain energy function with the additive form
\begin{align}
\Omega(\mathbf{F}, \mathbf{E}^e) = \Omega^m (\mathbf{F}) + \Omega^{\rm em}(\mathbf{F}, \mathbf{E}^e)  + \Omega^e( \mathbf{E}^e),
\end{align}
with $\Omega^m (\mathbf{F})$ referring to the {purely} mechanical behavior, $\Omega^{\rm em}(\mathbf{F}, \mathbf{E}^e)$ referring to the electomechanical coupling and $\Omega^e( \mathbf{E}^e)$ referring to the pure electric behavior. Accordingly, the first Piola-Kirchhoff stress can be written as two parts
\begin{align}
\mathbf{P}=\rho_0\frac{\partial \Omega^m}{\partial \mathbf{F}}+\rho_0\frac{\partial \Omega^{em}}{\partial \mathbf{F}}.
\end{align}

\section{Governing equations for electromechanical coupling in geometrically exact beam}
\begin{figure}[htb!]
	\centering
	\includegraphics[width=.6\textwidth]{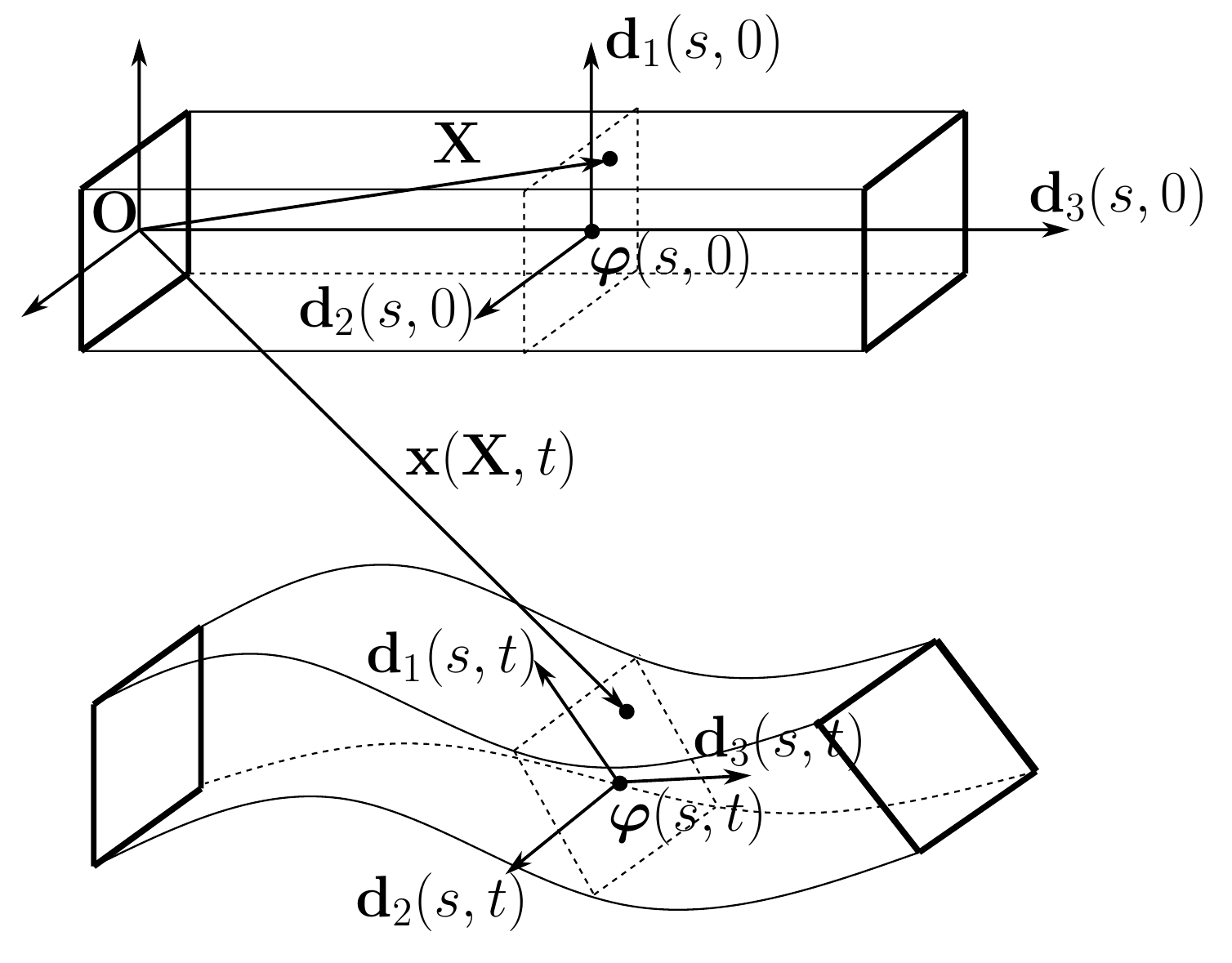}
	\caption{Configurations of the beam.}
	\label{fig1}
\end{figure}

In this work, the formulation of the electromechanical coupling problem presented above is extended to the geometrically exact beam. The deformation state of an initially straight beam over time can be distinguished by the initial configuration and the current configuration, as shown in Fig. \ref{fig1}.\\

In the Cosserat formulation of geometrically exact beam, the placement of a material point in the current configuration of the beam is given by
\begin{align}
\mathbf{x}(X^k,s,t)=\boldsymbol{\varphi}(s,t)+X^k \mathbf{d}_k(s,t), \;\;\;\; k=1,2. \label{x}
\end{align}
where $s \in [0,L] \subset \mathbb{R}$ denotes the arc-length of the line of the centroids $\boldsymbol{\varphi}(s,0)\in \mathbb{R}^3$ in the initial configuration, $\mathbf{d}_i(s,t)$ is an orthonormal triad at $s$ with the directors $\mathbf{d}_k(s,t), k=1,2$ spanning a principle basis of the cross section, and $X^k$ are the curvilinear coordinates on {the} cross section. Based on these assumptions, the governing equations for electromechanically coupled beam model can be consistently derived from the equations of continuum electromechanics.

\subsection{{Electromechanically coupled beam model in variational setting}}
{The governing equations of the beam are usually derived by integrating Eq. (\ref{blm}) in continuum mechanics over the beam cross section, see e.g. \citep{simo1985finite}. However, the Neumann boundary condition on the beam is not clearly described. To obtain a consistent derivation of beam equations from continuum mechanics considering the Neumann boundary conditions in Eq. (\ref{PNT}) and (\ref{DNQ}), in this work, we formulate the governing equations for the beam from the variational setting.}

The variational setting is formulated according to the Lagrange-d'Alembert principle
\begin{align}
\delta S+ \int_{0}^{T} \delta W^{\rm ext}dt=0, \label{dAp}
\end{align}
with $S$ the action and $W^{\rm ext}$ the external non-conservative work contributed by the body force and surface traction. According to the constitutive law in Eq. (\ref{CE}), the variation of the action can be formulated for the beam as
\begin{align}
\delta S&=\int_{0}^{T} \int_{B_0} \left(  \rho_0 \dot{\mathbf{x}} \cdot \delta \dot{\mathbf{x}} - \partial_{\mathbf{F}} \Omega : \delta \mathbf{F} -  \partial_{\mathbf{E}^e} \Omega \cdot \delta \mathbf{E}^e  \right) dVdt \nonumber\\
&=\int_{0}^{T} \left\lbrace \int_{c}\int_{\Sigma} \left[ \delta \mathbf{x} \cdot \left(   - \rho_0 \ddot{\mathbf{x}} + \nabla_{\mathbf{X}} \cdot \mathbf{P} \right)+ \delta \phi \left( - \nabla_{\mathbf{X}} \cdot \mathbf{D} \right) \right]  dAds +\int_{c} \oint_{\partial \Sigma} \left[ \delta \mathbf{x} \cdot \left( - \mathbf{P} \cdot \mathbf{N} \right) +\delta \phi \left( \mathbf{D}\cdot \mathbf{N} \right) \right] dlds \right\rbrace  dt, \label{dS}
\end{align}
where the formula $-\partial_{\mathbf{F}} \Omega : \delta \mathbf{F}= (\nabla_{\mathbf{X}} \cdot \partial_{\mathbf{F}} \Omega) \cdot \delta\mathbf{x} - \nabla_{\mathbf{X}} \cdot (\partial_{\mathbf{F}} \Omega \cdot \delta\mathbf{x})$ and the divergence theorem have been applied. Due to the assumption of geometrically exact beam, the volume integral in Eq. (\ref{dS}) is split into the curve integral over beam center line and the area integral over beam cross section $\Sigma$. At the same time, the surface integral in Eq. (\ref{dS}) is split into the curve integral over beam center line and the curve integral over the lateral contour of beam cross section. By using of $\mathbf{P}=\mathbf{t}_k\otimes\mathbf{d}_k(s,0)+\mathbf{t}_s\otimes\mathbf{d}_3(s,0)$  and $\mathbf{D}=D_k \mathbf{d}_k(s,0) + D_s\mathbf{d}_3(s,0)$, the force $\mathbf{f}$, the torque $\mathbf{m}$ and the electric displacement $d^e$ for the beam are defined as
 \begin{align}
\mathbf{f}=\int_{\Sigma}\mathbf{t}_s dA,\qquad \mathbf{m}=\int_{\Sigma} \mathbf{r} \times \mathbf{t}_s dA, \qquad d^e=\int_{\Sigma}D_s dA,
\end{align}
with $\mathbf{r}=\mathbf{x}-\boldsymbol{\varphi}$. For the sake of later use, the electric displacement vector $\mathbf{d}^e$ for the beam is defined as $\mathbf{d}^e=\begin{bmatrix} d_1^e & d_2^e & d_s^e \end{bmatrix}^T$ with the components $d_k^e=\int_{\Sigma} D_k dA$. In this case, the divergence terms in Eq. (\ref{dS}) can be formulated as
 \begin{align}
\int_{\Sigma} \nabla_{\mathbf{X}} \cdot \mathbf{P} dA&= \underbrace{\int_{\Sigma} \frac{\partial \mathbf{t}_1}{\partial X_1} +\frac{\partial \mathbf{t}_2}{\partial X_2} dA}_{\mathbf{f}^n} + \frac{\partial \mathbf{f}}{\partial s},\\
\int_{\Sigma} \mathbf{r} \times \nabla_{\mathbf{X}} \cdot \mathbf{P} dA&= \underbrace{\int_{\Sigma} \mathbf{r} \times\left( \frac{\partial \mathbf{t}_1}{\partial X_1} +\frac{\partial \mathbf{t}_2}{\partial X_2}\right)  dA}_{\mathbf{m}^n} + \frac{\partial \mathbf{m}}{\partial s}+\frac{\partial \boldsymbol{\varphi}}{\partial s} \times \mathbf{f},\\
\int_{\Sigma} \nabla \cdot \mathbf{D} dA&=\underbrace{ \int_{\Sigma} \frac{\partial D_1}{\partial X_1} +\frac{\partial D_2}{\partial X_2} dA}_{d^{en}} + \frac{\partial d^e}{\partial s}. \label{fd}
\end{align}
In the surface integral of Eq. (\ref{dS}), the unit normal vector is in the plane of cross section, i.e. $\mathbf{N}=n_k\mathbf{d}_k(s,0) (k=1,2)$. Thus, by applying the divergence theorem, we have $\oint_{\partial \Sigma} \mathbf{P}\cdot \mathbf{N}dl=\mathbf{f}^n$, $\oint_{\partial \Sigma}\mathbf{r} \times (\mathbf{P}\cdot \mathbf{N})dl=\mathbf{m}^n$ and $ \oint_{\partial \Sigma} \mathbf{D}\cdot \mathbf{N}dl=d^{en} $.\\
The variation of the position field is given by
\begin{align}
\delta \mathbf{x} = \delta \boldsymbol{\varphi} + \delta \boldsymbol{\eta} \times \mathbf{r},
\end{align}
with $\delta \boldsymbol{\eta}= \frac{1}{2}\mathbf{d}^i \times \delta \mathbf{d}_i$, see the details in \citep{eugster2014director}. The time derivatives of the position field are given by
\begin{align}
\dot{\mathbf{x}} = \dot{\boldsymbol{\varphi}} + \boldsymbol{\omega} \times \mathbf{r}, \qquad \ddot{\mathbf{x}} = \ddot{\boldsymbol{\varphi}} + \dot{\boldsymbol{\omega}} \times \mathbf{r} + \boldsymbol{\omega} \times \boldsymbol{\omega} \times \mathbf{r}, \label{dx} 
\end{align}
with $\boldsymbol{\omega}$ the spatial angular velocity. Using Eq. (\ref{fd}) - (\ref{dx}), the varitation of the action in Eq. (\ref{dS}) reads
\begin{align}
&\delta S=\nonumber\\
&\int_{c} \left[ \delta \boldsymbol{\varphi} \cdot \left(   - A_{\rho} \ddot{\boldsymbol{\varphi}} +\mathbf{f}^n+ \frac{\partial \mathbf{f}}{\partial s}  \right) + \delta \boldsymbol{\eta} \cdot \left(- \dot{\boldsymbol{\omega}} \mathbb{I} - \boldsymbol{\omega} \times \boldsymbol{\omega} \mathbb{I}+  \mathbf{m}^n+ \frac{\partial \mathbf{m}}{\partial s} + \frac{\partial \boldsymbol{\varphi}}{\partial s} \times \mathbf{f} \right) +\delta \phi \left(-d^{en}-\frac{\partial d^e}{\partial s} \right)  \right] ds\nonumber\\
& + \int_{c} \left[ \delta \boldsymbol{\varphi} \cdot \left( -\mathbf{f}^n\right) + \delta \boldsymbol{\varphi} \cdot \left( -\mathbf{m}^n\right) + \delta\phi \, d^{en} \right] ds, \label{ds2}
\end{align}
in which  $\mathbb{I}=\int_{\Sigma} \rho \left( \left\|  \mathbf{x}\right\|^2 - \mathbf{x} \otimes \mathbf{x} \right)  dA$ is the spatial mass moment of inertia tensor. It can be observed that $\mathbf{f}^n$  and $\mathbf{m}^n$ in the first curve integral of Eq. (\ref{ds2}) come from the divergence term of the action in Eq. (\ref{dS}), $\mathbf{f}^n$ and $\mathbf{m}^n$ in the second curve integral of Eq. (\ref{ds2}) come from the integration over a boundary surface in Eq. (\ref{dS}). Due to their opposite signs, they will be offset respectively.\\
By considering the Neumann boundary conditions in Eq. (\ref{PNT}) and (\ref{DNQ}), the variation of the external work is written as
\begin{align}
 \delta W^{\rm ext}&=\int_{B_0} \delta \mathbf{x} \cdot \rho_0\bar{\mathbf{b}} dV +\int_{\partial B_0} \delta \mathbf{x} \cdot \bar{\mathbf{T}} dA+ \int_{\partial B_0} \delta \phi \, \bar{Q} dA \nonumber\\
 &=\int_{c} \delta \boldsymbol{\varphi} \cdot \bar{\mathbf{f}} ds  + \int_{c}\delta \boldsymbol{\varphi} \cdot \bar{\mathbf{t}} ds + \int_{c} \delta \boldsymbol{\eta} \cdot \bar{\mathbf{m}} ds  + \int_{c}\delta \boldsymbol{\eta} \cdot \bar{\boldsymbol{\tau}} ds +\int_{c} \delta \phi \, \bar{q} ds, \label{wext}
\end{align}
where the prescribed body force, lateral traction, torque and lateral charge are defined respectively as
\begin{align}
\bar{\mathbf{f}}=\int_{\Sigma}\rho_0 \bar{\mathbf{b}}dA, \quad \bar{\mathbf{t}}= \oint_{\partial \Sigma} \bar{\mathbf{T}} dl, \quad \bar{\mathbf{m}}=\int_{\Sigma}\mathbf{r}\times \rho_0 \bar{\mathbf{b}}dA, \quad \bar{\boldsymbol{\tau}}= \oint_{\partial \Sigma} \mathbf{r} \times \bar{\mathbf{T}} dl, \quad \bar{q}= \oint_{\partial \Sigma} \bar{Q} dl.
\end{align} 
Combing Eq. (\ref{ds2}) and (\ref{wext}), the Lagrange-d'Alembert principle for the beam is given by
\begin{align}
\int_{0}^{T} \left\lbrace \int_{c} \left[ \delta \boldsymbol{\varphi} \cdot \left(   - A_{\rho} \ddot{\boldsymbol{\varphi}} + \frac{\partial \mathbf{f}}{\partial s} +\bar{\mathbf{f}} +\bar{\mathbf{t}}  \right) +\delta \boldsymbol{\eta} \cdot \left(- \dot{\boldsymbol{\omega}} \mathbb{I} - \boldsymbol{\omega} \times \boldsymbol{\omega} \mathbb{I}+  \frac{\partial \mathbf{m}}{\partial s} + \frac{\partial \boldsymbol{\varphi}}{\partial s} \times \mathbf{f} +\bar{\mathbf{m}} +\bar{\boldsymbol{\tau}} \right) \right. \right. \nonumber\\
\left. \left. +\delta \phi \left(-\frac{\partial d^e}{\partial s} +\bar{q} \right)  \right] ds \right\rbrace  dt=0\label{dsdw}
\end{align}

\subsection{Balance of linear and angular momentum}
The requirement of stationary in Eq. (\ref{dsdw}) in terms of the translation field $ \boldsymbol{\varphi}$ leads to the balance of linear momentum for the beam
\begin{align}
\frac{\partial \mathbf{f}}{\partial s}+ \bar{\mathbf{f}}+ \bar{\mathbf{t}} =A_{\rho} \ddot{\boldsymbol{\varphi}} \qquad s \in [0,L]. \label{blm-b}
\end{align}
The requirement of stationary in Eq. (\ref{dsdw}) in terms of the rotation field $ \boldsymbol{\eta}$ leads to the balance of the angular momentum {for the} beam
\begin{align}
 \frac{\partial \mathbf{m}}{\partial s} + \frac{\partial \boldsymbol{\varphi}}{\partial s} \times \mathbf{f} +\bar{\mathbf{m}} +\bar{\boldsymbol{\tau}} =\mathbb{I} \dot{\boldsymbol{\omega}} +\boldsymbol{\omega} \times \mathbb{I} \boldsymbol{\omega} \qquad s \in [0,L]. \label{bam-b}
\end{align}

\subsection{Maxwell equation}
The requirement of stationary in Eq. (\ref{dsdw}) in terms of the electric potential $ \phi$ leads to the Maxwell equation for the beam
\begin{align}
\partial_s d_s^e - \bar{q}  =0 \qquad s \in [0,L]. \label{mwb}
\end{align}
\subsection{Electromechanical coupling}
The force $\mathbf{f}$, the torque $\mathbf{m}$ and the electric displacement vector $\mathbf{d}^e$ in the beam governing equations are related with the kinematic variables via the constitutive equations
\begin{align}
\mathbf{d}^e=-\rho_0\frac{\partial \Omega_b( \boldsymbol{\gamma}, \boldsymbol{\kappa}, \boldsymbol\varepsilon)}{\partial \boldsymbol\varepsilon},  \;\;\;\;  \mathbf{f}=\rho_0\frac{\partial \Omega_b( \boldsymbol{\gamma}, \boldsymbol{\kappa}, \boldsymbol\varepsilon)}{\partial \boldsymbol{\gamma}},  \;\;\;\;  \mathbf{m}=\rho_0\frac{\partial \Omega_b( \boldsymbol{\gamma}, \boldsymbol{\kappa}, \boldsymbol\varepsilon)}{\partial \boldsymbol{\kappa}}, \label{dfm}
\end{align}
where $\Omega_b$ is the strain energy per unit arc-length in {a} beam, $\boldsymbol\varepsilon$ is the strain-like electrical variable conjugated with the electric displacement $\mathbf{d}^e$ of {the} beam, $\boldsymbol{\gamma}$ and $\boldsymbol{\kappa}$ are the  beam strain measures conjugated with $\mathbf{f}$ and $\mathbf{m}$, respectively. The electromechanical coupling can be specified by the strain energy function in the additive form
\begin{align}
\Omega_b( \boldsymbol{\gamma}, \boldsymbol{\kappa}, \boldsymbol\varepsilon) = \Omega^{m}_b( \boldsymbol{\gamma}, \boldsymbol{\kappa})  + \Omega^{\rm em}_b( \boldsymbol{\gamma}, \boldsymbol{\kappa}, \boldsymbol\varepsilon)  + \Omega^{e}_b(\boldsymbol\varepsilon).
\end{align}
Due to the coupled term in strain energy, the force and torque are contributed by two parts accordingly, the electric part and the mechanical part, i.e.
\begin{align}
\mathbf{f}=\rho_0\frac{\partial \Omega^m_b}{\partial \boldsymbol{\gamma}}+\rho_0\frac{\partial \Omega^{em}_b}{\partial \boldsymbol{\gamma}},\;\;\;\;\; \mathbf{m}=\rho_0\frac{\partial \Omega^m_b}{\partial \boldsymbol{\kappa}}+\rho_0\frac{\partial \Omega^{em}_b}{\partial \boldsymbol{\kappa}}.
\end{align}
\section{Mechanical and electrical kinematics in {the} beam}
To compute the force $\mathbf{f}$, the torque $\mathbf{m}$ and the electric displacement $\mathbf{d}^e$ in the governing equations of the electromechanically coupled beam model in Section 3, the conjugated kinematic variables for mechanical and electrical parts have to be formulated. The mechanical strain measures $\boldsymbol{\gamma}$ and $\boldsymbol{\kappa}$ for the geometrically exact beam have been defined and related with the deformation gradient in the literature, see e.g. \citep{auricchio2008geometrically}. However, the electric potential $\phi$ as well as the strain-like electrical variable $\boldsymbol\varepsilon$ are still not given for the beam.

\subsection{Mechanical kinematics in {the} beam}
By setting the origin of the global Cartesian coordinate system $\mathbf{O}$ {to} one end of the beam as shown in Fig. \ref{fig1}, the location of the  node $\boldsymbol{\varphi}(s,0)$ of the straight beam can be rewritten in terms of the director $\mathbf{d}_3(s,0)$ and the arc length $s$
\begin{align}
\boldsymbol{\varphi}(s,0)=X^3 \mathbf{d}_3(s,0)=s \mathbf{d}_3(s,0), \;\;\;\; {\rm or } \;\;\; X^3 =s.
\end{align}
Thus the components $X^i$ of the vector $\mathbf{X}$ in the initial configuration can be computed by projecting the position vector to the directors
\begin{align}
X^i=\mathbf{X}\cdot \mathbf{d}_i(s,0),\;\;\;\; i=1,2,3.
\label{Xi}
\end{align} 

By applying Eq. (\ref{x}) and Eq. (\ref{Xi}), the deformation gradient at a point $(X^1, X^2, s)$ in {the} beam can be written as, see \citep{auricchio2008geometrically},
\begin{align}
\mathbf{F}(X^1, X^2, s,t)&=\frac{\partial \mathbf{x}}{\partial \mathbf{X}}=\frac{\partial \mathbf{x}}{\partial X_i} \otimes  \mathbf{d}_i(s,0) \nonumber\\
&=\left[ \mathbf{I} + \left(\frac{\partial \boldsymbol{\varphi}(s,t)}{\partial s} - \mathbf{d}_3(s,t) + X^1 \frac{\partial \mathbf{d}_1(s,t)}{\partial s} + X^2 \frac{\partial \mathbf{d}_2(s,t)}{\partial s} \right)\otimes  \mathbf{d}_3(s,t) \right] \boldsymbol{\Lambda}(s),
\label{F}
\end{align}
with the rotation tensor $ \boldsymbol{\Lambda}(s)=\mathbf{d}_i(s,t) \otimes  \mathbf{d}_i(s,0)$ and $\boldsymbol{\Lambda}(s)^{-1}=\boldsymbol{\Lambda}(s)^T $. The derivatives in Eq. (\ref{F}) can be written in terms of the beam strain measures, such as $\frac{\partial \mathbf{d}_k(s,t)}{\partial s}$ is related to the strain $\boldsymbol{\kappa}$ for bending and torsion by
\begin{align} 
\frac{\partial \mathbf{d}_k(s,t)}{\partial s}=\boldsymbol{\kappa}(s,t) \times \mathbf{d}_k(s,t)=\left[ \kappa_j\mathbf{d}_j(s,t)\right]  \times \mathbf{d}_k(s,t),
\end{align} 
and $\frac{\partial \boldsymbol{\varphi}(s,t)}{\partial s}$ is related to the strain $\boldsymbol{\gamma}$ for shear and elongation by
\begin{align} 
\frac{\partial \boldsymbol{\varphi}(s,t)}{\partial s} - \mathbf{d}_3(s,t)=\boldsymbol{\gamma}(s,t).
\end{align} 
{Consequently}, the deformation gradient $\mathbf{F}$ in Eq. (\ref{F}) can be further formulated in terms of the beam strain measures as
\begin{align} 
\mathbf{F}=\left\lbrace  \mathbf{I} + \left[ \boldsymbol{\gamma}(s,t) + \boldsymbol{\kappa}(s,t) \times X^k \mathbf{d}_k(s,t)\right]\otimes  \mathbf{d}_3(s,t)\right\rbrace   \boldsymbol{\Lambda}(s)
=\left[ \mathbf{I} +\mathbf{a}(s,t) \otimes  \mathbf{d}_3(s,t) \right]  \boldsymbol{\Lambda}(s)
\end{align}
with $\mathbf{a}(s,t) =  \boldsymbol{\gamma}(s,t) + \boldsymbol{\kappa}(s,t) \times X^k \mathbf{d}_k(s,t) $. The variable $\mathbf{a}$ can be formulated in the reference configuration with the beam strain measures $\boldsymbol{\Gamma}$ and $\mathbf{K}$
\begin{align} 
\mathbf{a}^r=\boldsymbol{\Lambda}^T\mathbf{a}=\boldsymbol{\Gamma} + \mathbf{K} \times X^k  \mathbf{d}^0_k. \label{ar}
\end{align}
The determinant of {the} deformation gradient reads
\begin{align} 
J={\rm det}(\mathbf{F})= 1+\mathbf{a}(s,t) \cdot \mathbf{d}_3(s,t).
\end{align}
Accordingly, the Green-Lagrange strain and the right Cauchy Green tensor are given by
\begin{align} 
\mathbf{E}&=\left[ \mathbf{a}^r \otimes \mathbf{d}_3(s,0)\right]^{\rm sym} +  \frac{1}{2} (\mathbf{a}^r \cdot \mathbf{a}^r) \mathbf{d}_3(s,0) \otimes\mathbf{d}_3(s,0),\\
\mathbf{C}&=\mathbf{I} + 2 \left[ \mathbf{a}^r \otimes \mathbf{d}_3(s,0)\right]^{\rm sym} + (\mathbf{a}^r \cdot \mathbf{a}^r) \mathbf{d}_3(s,0) \otimes\mathbf{d}_3(s,0). \label{EC}
\end{align}
The inverse of the right Cauchy Green tensor can be formulated as
\begin{align}
\mathbf{C}^{-1}&=\mathbf{F}^{-1}\mathbf{F}^{-T}=\boldsymbol{\Lambda}^{-1}\mathbf{A}^{-1}\mathbf{A}^{-T}\boldsymbol{\Lambda}^{-T}=\boldsymbol{\Lambda}^{T}\mathbf{A}^{-1}\mathbf{A}^{-T}\boldsymbol{\Lambda}
\end{align}
with $\mathbf{A}^{-1}=\mathbf{I}-\frac{1}{J}\left[ \mathbf{a}\otimes  \mathbf{d}_3(s,t)\right] $. 

The material time derivative of the location of a material point in the current configuration reads
\begin{align}
\dot{\mathbf{x}}(X^k,s,t)=\dot{\boldsymbol{\varphi}}(s,t)+X^k \dot{\mathbf{d}}_k(s,t),
\label{dotx}
\end{align}
by which the time derivative of deformation gradient can be evaluated with
\begin{align}
&\dot{\mathbf{F}}(X^1, X^2, s,t)=\frac{\partial \dot{\mathbf{x}}}{\partial \mathbf{X}}=\frac{\partial \dot{\mathbf{x}}}{\partial X_i} \otimes  \mathbf{d}_i(s,0) \nonumber\\
&=\dot{\mathbf{d}}_1(s,t) \otimes  \mathbf{d}_1(s,0) + \dot{\mathbf{d}}_2(s,t) \otimes  \mathbf{d}_2(s,0) + \left( \frac{\partial \dot{\boldsymbol{\varphi}}(s,t)}{\partial s} + X^1 \frac{\partial \dot{\mathbf{d}}_1(s,t)}{\partial s} + X^2 \frac{\partial \dot{\mathbf{d}}_2(s,t)}{\partial s}  \right)  \otimes  \mathbf{d}_3(s,0) \nonumber\\
&= \dot{\boldsymbol{\Lambda}}(s) + \left( \frac{\partial \dot{\boldsymbol{\varphi}}(s,t)}{\partial s} - \dot{\mathbf{d}}_3(s,t) + X^1 \frac{\partial \dot{\mathbf{d}}_1(s,t)}{\partial s} + X^2 \frac{\partial \dot{\mathbf{d}}_2(s,t)}{\partial s} \right) \otimes  \mathbf{d}_3(s,0).
\label{dotF}
\end{align}

\subsection{Electrical kinematics in {the} beam}
\begin{figure}[htb!]
	\centering
	\includegraphics[width=.8\textwidth]{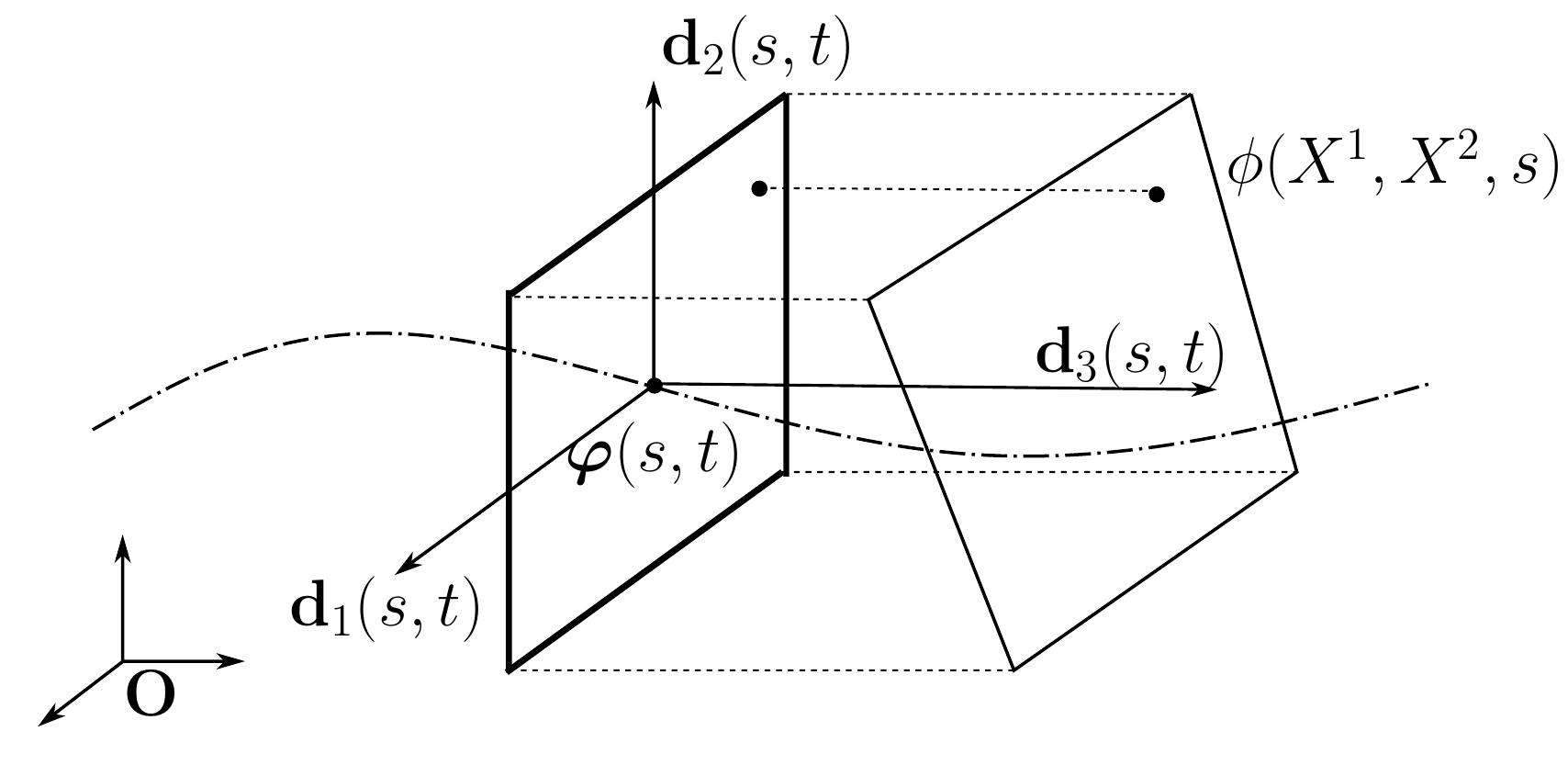}
	\caption{Linear distribution of electric potential on cross section.}
	\label{beam_e}
\end{figure}

To {formulate} the electromechanical coupling problem, the electric potential will {serve} as the extra degree of freedom. In this work, the electric potential at the point $(X^1,X^2,s)$ is represented by the electric potential at {the} beam node plus the increment from the beam node to the point $(X^1,X^2)$ on the cross section as shown in Fig. \ref{beam_e}. Similar to the local description of the cross section in Eq. (\ref{x}), the electric potential on {the} cross section is given by
\begin{align} 
\phi (X^1,X^2,s)=\phi _o(s) + X^1 \alpha(s) + X^2 \beta(s)
\label{linearV}
\end{align}
with $\phi _o(s) $ the electric potential at the beam node, $ \alpha(s)$ and $\beta(s)$ the incremental parameters of the electric potential in the directions of $\mathbf{d}_1$ and $\mathbf{d}_2$, respectively.

The  Eq. (\ref{x}) describes the cross section as a plane. Correspondingly, Eq. (\ref{linearV}) defines a linear distribution of electric potential on the cross section. If the electric potential is constant within the cross section (i.e. $\alpha,\beta=0$), the electric field only exists in the $\mathbf{d}_3$ direction, which will lead to the uniaxial contraction in beam.

According to the Maxwell equations, the electric field is defined as the gradient of the  electric potential $ \phi$, see Eq. (\ref{E}). To compute the gradient of the electric potential for {the} beam, {a} similar approach as the deformation gradient in Eq. (\ref{F}) can be adopted. By applying the electric potential in Eq. (\ref{linearV}), the electric field at $(X^1,X^2,s)$ in {the} beam can be computed as
\begin{align} 
\mathbf{E}^e(X^1,X^2,s)&= -\frac{\partial \phi}{\partial X_i} \otimes  \mathbf{d}_i(s,0)\\
&=-\left[ \alpha(s)  \mathbf{d}_1(s,0) + \beta(s)  \mathbf{d}_2(s,0) + \left( \frac{\partial \phi_o(s)}{\partial s}  + X^1 \frac{\partial  \alpha(s)}{\partial s} + X^2 \frac{\partial  \beta(s)}{\partial s} \right) \mathbf{d}_3(s,0) \right]. \label{Ee}
\end{align}
For the later computing of the virtual work induced by the electric field, the variation of the electric field is given by
\begin{align} 
\delta \mathbf{E}^e = -\left[ \delta \alpha(s)  \mathbf{d}_1(s,0) + \delta \beta(s)  \mathbf{d}_2(s,0) + \delta\left( \frac{\partial \phi_o(s)}{\partial s}  + X^1 \frac{\partial  \alpha(s)}{\partial s} + X^2 \frac{\partial  \beta(s)}{\partial s} \right) \mathbf{d}_3(s,0) \right]. \label{dE}
\end{align}
Since the electric field expressed in Eq. (\ref{Ee}) is not a strain-like variable for beam, we need to formulate the strain-like electrical variable $\boldsymbol{\varepsilon}$, which can be consistently conjugated with the beam electric displacement $\mathbf{d}^e$ in Eq. (\ref{dfm}). For this purpose, we start from the internal virtual work $\delta W_{\rm int}$ induced by the electric field in the 3D continuum
\begin{align} 
\delta W_{\rm int}=\int_{B_0} \mathbf{D}\cdot \delta \mathbf{E}^e dV.
\end{align}
By applying the electric displacement in Eq. (\ref{mwb}) and the variation of electric field in Eq. (\ref{dE}) to the internal virtual work, we obtain
\begin{align} 
\delta W_{int}&= \int_{B_0} (D_i \mathbf{d}_i) \cdot  \left[ \delta \alpha(s)  \mathbf{d}_1(s,0) + \delta \beta(s)  \mathbf{d}_2(s,0) + \delta\left( \frac{\partial \phi_o(s)}{\partial s}  + X^1 \frac{\partial  \alpha(s)}{\partial s} + X^2 \frac{\partial  \beta(s)}{\partial s} \right) \mathbf{d}_3(s,0) \right]  dV \nonumber\\
&=\int_c  \left[ \delta \alpha(s) \int_{\Sigma} D_1 dA +\delta \beta(s)  \int_{\Sigma} D_2 dA + \delta \frac{\partial \phi_o(s)}{\partial s} \int_{\Sigma} D_3 dA   \right]ds \nonumber\\
&= \int_c  \mathbf{d}^e \cdot \delta \boldsymbol{\varepsilon} ds \label{Wint},
\end{align}
where the work conjugate of $\mathbf{d}^e$ and $\boldsymbol{\varepsilon} $ can be observed with the strain-like electrical variable $\boldsymbol{\varepsilon}$ defined as
\begin{align} 
\boldsymbol{\varepsilon}=
\begin{bmatrix}
\alpha(s) &  \beta(s) &  \frac{\partial \phi_o(s)}{\partial s}
\end{bmatrix}^T. \label{estrain}
\end{align}

\section{Strain energy function for {the} beam}
The strain energy function of {the} beam model has been derived for the hyperelastic constitutive law, such as the widely used Saint-Venant-Kirchhoff model \citep{simo1986three}. However, the strain energy for other material behaviors are not given for {the} beam model, including the dielectric elasticity.  In continuum electromechanics, many material models have been postulated to describe different material behaviors, such as the  dielectric elastomers in \citep{vu2007numerical}. To apply these material models in {a} beam formulation, the rewriting of the strain energy function in terms of the beam strains is required, which can be achieved by applying the kinematic relations introduced in Section 4. In this part, we firstly present the derivation of the widely used Saint-Venant-Kirchhoff model for {the} beam. Then, the strain energy function of the dielectric elastomer is formulated for {the} beam such that it can be used in {the} beam constitutive equations directly. 
\subsection{Saint-Venant-Kirchhoff model}
The Saint-Venant-Kirchhoff model describes a linear relationship between the Green-Lagrange strain $\mathbf{E}$ and the second Piola-Kirchhoff stress $\mathbf{S}$ with the strain energy function
\begin{align}
\Omega (\mathbf{E}) =\frac{1}{2}\lambda \left({\rm tr}\mathbf{E}\right)^2 + \mu \mathbf{E} : \mathbf{E},
\end{align}
in which $\lambda$ and $\mu$ are the Lam{\'e} parameters. Inserting the linear part of $\mathbf{E}$ in Eq. (\ref{EC}) into the previous equation, the reduced strain energy $\Omega^r$ reads
\begin{align} 
\Omega^r=\frac{1}{2} \left\lbrace  \lambda \left[ {\rm tr}(\mathbf{a}^r \otimes \mathbf{d}_3(s,0))^{\rm sym}\right]^2  + 2\mu \left[ \frac{1}{2} (\mathbf{a}^r \cdot \mathbf{d}_3(s,0))^2 +\frac{1}{2}  \mathbf{a}^r \cdot \mathbf{a}^r  \right] \right\rbrace  = \frac{1}{2} \mathbf{a}^{rT} \mathbf{D} \mathbf{a}^{r}
\end{align}
with $\mathbf{D}=( \lambda  + \mu)\mathbf{d}_3(s,0) \otimes \mathbf{d}_3(s,0)  + \mu \mathbf{I}   $. Integrating $\Omega^r$ over beam cross section, the reduced strain energy for {the} beam is obtained as
\begin{align} 
\Omega^{r}_b (\boldsymbol{\Gamma}, \mathbf{K})&=\int_{\Sigma}\Omega^r dA = \frac{1}{2} \int_{\Sigma} (\boldsymbol{\Gamma} + \mathbf{K} \times \mathbf{r})^T  \mathbf{D} (\boldsymbol{\Gamma} + \mathbf{K} \times \mathbf{r}) dA \nonumber\\
&= \frac{1}{2}  \boldsymbol{\Gamma} ^T   \mathbf{D}^N \boldsymbol{\Gamma} + \frac{1}{2} \mathbf{K}^T \mathbf{D}^K \mathbf{K},  \label{SVT}
\end{align} 
where the material tangents $\mathbf{D}^N$ and $\mathbf{D}^K$ are given by
\begin{align}
\mathbf{D}^N=
\begin{bmatrix}
\mu A & 0 &0\\
0 & \mu A  &0\\
0 & 0 & (\lambda+2\mu) A 
\end{bmatrix}, \;\;\;\;\;
\mathbf{D}^K=
\begin{bmatrix}
(\lambda+2\mu) J_{22} & (\lambda+2\mu) J_{12} &0\\ 
(\lambda+2\mu) J_{21} & (\lambda+2\mu) J_{11} &0\\
0 & 0 & \mu J_p
\end{bmatrix}
\end{align}
with the first moments of area $J_{ii}=\int_{\Sigma} X_i X_i dA$, the product moments of area $J_{12}=J_{21}=-\int_{\Sigma} X_1 X_2 dA$ and the polar moment of area $J_p=\int_{\Sigma} X_1^2+ X_2^2 dA$.  It can be observed that the widely used beam constitutive model postulated by \citep{simo1986three} and \citep{simo1991geometrically} are recovered when $ \lambda+2\mu = E$. According to the formula of Lam{\'e} parameters,  {this formulation assumes the limit of Poisson's  $\nu \rightarrow 0$, i.e. no lateral contraction. However, the physics is incorrect and it leads to the incorrect material moduli. To overcome this problem, the uniaxial stress assumption can be additionally included within a mix formulation such as the Hellinger-Reissner principle. An alternative solution is to apply a very small Poisson's ratio $\nu$ in the simulation,} by which a relatively fair comparison between the beam model and the 3D finite element model can be achieved as shown in the numerical examples of this work.\\
Since the reduced beam strain energy $\Omega^{r}_b$ is derived by only inserting the linear part of the Green-Lagrange strain $\mathbf{E}$, the derived beam force will be just the linear part of the full one. To explain it, we start from the current force $\mathbf{n}$ of the beam computed from the first Piola-Kirchhoff stress $\mathbf{P}$
\begin{align}
\mathbf{n}=\int_{\Sigma} \mathbf{P} \mathbf{d}_3(s,0) dA=\int_{\Sigma} \mathbf{FS} \mathbf{d}_3(s,0) dA=\int_{\Sigma} \boldsymbol\Lambda \left[ \mathbf{I} + \mathbf{a}^r \otimes \mathbf{d}_3(s,0)\right] \mathbf{S} \mathbf{d}_3(s,0) dA.
\end{align}
By applying the relation $\mathbf{n}=\boldsymbol{\Lambda}\mathbf{N}$, the reference beam force $\mathbf{N}$ reads
\begin{align}
\mathbf{N}=\int_{\Sigma}  \left[ \mathbf{I} + \mathbf{a}^r \otimes \mathbf{d}_3(s,0)\right] \mathbf{S}\mathbf{d}_3(s,0) dA. \label{Nf}
\end{align}
It can be observed that force $\mathbf{N}$ is composed of two parts, one depending only on the stress and another also depending on strain variable $\mathbf{a}^r$. It has been proven that $\mathbf{N}$ is work conjugated with the beam strain $\boldsymbol{\Gamma}$, see the details in \citep{auricchio2008geometrically}. When the reduced strain energy in Eq. (\ref{SVT}) is applied, the derived force $\mathbf{N}^r$ equals to the integration of the second Piola-Kirchhoff stress $\mathbf{S}^{\rm lin}$ computed with the linear strain $\mathbf{E}^{\rm lin}$
 \begin{align}
 \mathbf{N}^r=\frac{\partial \Omega^{r}_b}{\partial \boldsymbol{\Gamma}}= \int_{\Sigma} \mathbf{S}^{\rm lin}\mathbf{d}_3(s,0)dA,
 \end{align}
which can be seen as the linear part of the full force $\mathbf{N}$ in Eq. (\ref{Nf}).


\subsection{Extended Neo-Hookean model for DEA}
To model the DEA, the material model of the dielectric elasticity applied in \citep{schlogl2016electrostatic} for the finite element simulation is applied to the beam model in this work, where the strain energy density is given by
\begin{align}
\Omega(\mathbf{C},\mathbf{E}^e)=\underbrace{ \frac{\mu}{2} \left( \mathbf{C}:\mathbf{1}-3 \right) - \mu {\rm ln} J + \frac{\lambda}{2} ({\rm ln} J)^2}_{\text{Neo-Hookean}} +\underbrace{c_1 \mathbf{E}^e \cdot \mathbf{E}^e + c_2 \mathbf{C}:(\mathbf{E}^e \otimes \mathbf{E}^e)}_{\text{Polarization in dielectric material}} - \underbrace{ \frac{1}{2} \varepsilon_0 J \mathbf{C}^{-1} : (\mathbf{E}^e \otimes \mathbf{E}^e)}_{\text{Free space term in vacuum}} \label{omega}
\end{align}
with $\varepsilon_0$ the vacuum permittivity, $c_1$ and $c_2$ the electrical parameters. It can be observed that the strain energy is composed of three parts, the Neo-Hookean part referring to the pure elastic behavior, the polarization part referring to the polarization in the condensed matter and the free space part referring to the effect in vacuum. The last two terms in Eq. (\ref{omega}) characterize the electromechanical coupling.

Apart from the dielectric elasticity, the viscoelastic effect in the dielectric material is accounted for by means of the first Piola-Kirchhoff stress $\mathbf{P}^{\rm vis}$, see the Kelvin-Voigt model in \citep{wriggers2008nonlinear},
\begin{align}
\mathbf{P}^{\rm vis}=\frac{1}{2}J\eta \left( \mathbf{F}^{-T}\cdot \dot{\mathbf{F}}^T \cdot \mathbf{F}^{-T} + \dot{\mathbf{F}}\cdot \mathbf{C}^{-1}\right), \label{Pv}
\end{align}
in which $\eta$ is the damping parameter. 

The strain energy  function for {the} beam corresponding to the continuum model in Eq. (\ref{omega}) can be derived with the same procedure as the Saint-Venant-Kirchhoff model. By neglecting the free space term in Eq. (\ref{omega}), the strain energy function for beam is obtained by integrating $\Omega(\mathbf{C},\mathbf{E}^e)$ over the cross section
\begin{align}
\Omega_b (\boldsymbol{\Gamma}, \mathbf{K}, \boldsymbol{\varepsilon}) =\int_{\Sigma} \Omega(\mathbf{C},\mathbf{E}^e) dA, \label{Ob}
\end{align}
{where the integration can be evaluated with the numerical approach as well as the analytical approach. As the analytical approach, the beam strain energy function $\Omega_b$ is explicitly formulated in the Appendix.}

\section{Discrete variational integration scheme with null space projection}
\subsection{Discrete Euler–Lagrange equations}
In this work, the electromechanically coupled beam dynamics is {approximated} within the constrained discrete variational scheme with the null space projection. The Lagrange-d'Alembert principle for the constrained system reads
\begin{align}
\delta \int_{0}^{T} \left[  L( \mathbf{q}, \dot{\mathbf{q}}) - \mathbf{g}^T(\mathbf{q})\cdot \boldsymbol\lambda \right] dt + \int_{0}^{T}\mathbf{f}^{\rm ext}( t) \cdot \delta \mathbf{q}dt=0,
\end{align}
where  $\mathbf{q}$ is the configuration, $L( \mathbf{q}, \dot{\mathbf{q}})$ is the Lagrangian, $\mathbf{g}$ {represents holonomic} constraints, $\boldsymbol\lambda$ is the Lagrangian multiplier and $\mathbf{f}^{\rm ext}(t)$ is the external force. By considering the electrical effect in geometrically exact beam, the electric potential $\phi_o$ and the incremental variables $(\alpha, \beta )$ in Eq. (\ref{linearV}) are treated as the electrical degrees of freedom $\boldsymbol\phi=\begin{bmatrix}  \phi_o& \alpha& \beta \end{bmatrix}$ such that the configuration of the beam model is extended to
\begin{align}
\mathbf{q}=\begin{bmatrix}
\boldsymbol{\varphi}& \mathbf{d}_1
&\mathbf{d}_2&\mathbf{d}_3& \boldsymbol\phi
\end{bmatrix}^T. \label{q}
\end{align}
According to the {kinematic} assumptions in geometrically exact beams, the directors have to fulfill the orthogonal constraints
\begin{align}
\mathbf{g}( \mathbf{q})=\begin{bmatrix}                           
\frac{1}{2}(\mathbf{d}_1^T \mathbf{d}_1-1) \\                                               
\frac{1}{2}(\mathbf{d}_2^T \mathbf{d}_2-1)\\   
\frac{1}{2}(\mathbf{d}_3^T \mathbf{d}_3-1) \\  
\mathbf{d}_1^T\mathbf{d}_2\\
\mathbf{d}_1^T\mathbf{d}_3\\ 
\mathbf{d}_2^T\mathbf{d}_3
\end{bmatrix}=\mathbf{0}.
\end{align}
The continuous Lagrangian contains the difference between the kinetic energy $T(\dot{\mathbf{q}}) $ and the internal potential energy $V(\mathbf{q})$
\begin{align}
L(\mathbf{q}, \dot{\mathbf{q}})= T(\dot{\mathbf{q}}) -V(\mathbf{q}).
\end{align}
Since the electrical variables do not contribute to the kinetic energy, the kinetic energy for geometrically exact beams is computed as
\begin{align}
T=  \int_c \left( \frac{1}{2} A_{\rho} \left| \dot{\boldsymbol  \varphi} \right| ^2 + \frac{1}{2} \sum_{i=1}^{2} M^i_{\rho}\left| \dot{\mathbf{d}}_i \right| ^2 \right) ds, \label{T}
\end{align}
where $A_{\rho}$ is the mass density per reference arc-length and $M^i_{\rho}$ are the principle mass moments of inertia of cross section.   In accordance with the configuration defined in Eq. (\ref{q}), the component of the consistent mass matrix corresponding to the electrical degree of freedom $\boldsymbol\phi$ will be zero.

For the coupled hyperelastic material in DEA, the internal potential energy is computed  by an integration of the beam strain energy density $ \Omega_b$ in Eq. (\ref{Ob}) over the beam center line
\begin{align}
V(\mathbf{q}) = \int_c \Omega_b (s) ds.
\end{align}
The external force $\mathbf{f}^{\rm ext}$ contains all non-conservative forces, such as the viscoelastic effect in this work. Based on the Kelvin-Voigt model in Eq. (\ref{Pv}), the non-conservative work contributed by the viscoelastic effect is given by
\begin{align}
W^{\rm vis}=\int_{B_0}  \mathbf{P}^{\rm vis}: \mathbf{F}dV.
\end{align}
In this case, the external force corresponding to the viscoelastic effect can be formulated as
\begin{align}
\mathbf{f}^{\rm vis}(\mathbf{q},\dot{\mathbf{q}})=\frac{\partial W^{\rm vis}}{\partial \mathbf{q}}= \int_{B_0} \frac{\partial W^{\rm vis}}{\partial \mathbf{F} } : \frac{\partial \mathbf{F}}{\partial \mathbf{q} } dV =  \int_c \int_{\Sigma} \mathbf{P}^{\rm vis}: \frac{\partial \mathbf{F}}{\partial \mathbf{q} }dAds.
\end{align}

The beam is first spatially discretized with the 1D finite elements. Then the variational integration scheme, see e.g. \citep{leyendecker2008variational}, is applied to temporally discretize the action of the dynamic system, by which the {good} long term energy {behavior} can be obtained. In the variational integration scheme, the action integral within the time interval $(t_n,t_{n+1})$ is approximated with the discrete Lagrangian $L_d$ as
\begin{align}
\int_{t_n}^{t_{n+1}} L(\mathbf{q},  \dot{\mathbf{q}})dt \approx L_d(\mathbf{q}_n,\mathbf{q}_{n+1}) = \Delta t L(\frac{\mathbf{q}_{n+1}+\mathbf{q}_n}{2},\frac{\mathbf{q}_{n+1}-\mathbf{q}_n}{\Delta t}),
\end{align}
where the discrete Lagrangian $L_d$ is computed by applying the finite difference approximation to the velocity $\dot{\mathbf{q}}$ and the midpoint {rule} to the configuration $\mathbf{q}$, i.e.
\begin{align}
\dot{\mathbf{q}}\approx \frac{\mathbf{q}_{n+1}-\mathbf{q}_n}{\Delta t}, \;\;\;\;\; \mathbf{q}\approx\frac{\mathbf{q}_{n+1}+\mathbf{q}_n}{2} \label{mid}.
\end{align}
After the temporal discretisation, the discrete Euler–Lagrange equations can be obtained by taking the variation of the discrete action and requiring stationarity. To eliminate the constraint forces $\boldsymbol\lambda$ from the system, see e.g. \citep{betsch2006discrete}, the nodal reparametrisation $\mathbf{q}_{n+1} = \mathbf{F}_d (\mathbf{u}_{n+1}, \mathbf{q}_{n})$ and the discrete null space matrix $\mathbf{P}_d$ are applied to the discrete Euler–Lagrange equations leading to
\begin{align}
\mathbf{P}_d^T(\mathbf{q}_n) \left[ \frac{\partial L_d(\mathbf{q}_{n-1}, \mathbf{q}_{n})}{\partial \mathbf{q}_{n}} + \frac{\partial L_d\left( \mathbf{q}_{n}, \mathbf{F}_d (\mathbf{u}_{n+1}, \mathbf{q}_{n})\right) }{\partial \mathbf{q}_{n}} + \mathbf{f}_n^{\rm ext-} + \mathbf{f}_n^{\rm ext+} \right] = \mathbf{0}, \label{R}
\end{align}
where $\mathbf{u}_{n+1}$ is the generalized configuration acting as the unknown variable, $\mathbf{f}_n^{\rm ext-}$ and $\mathbf{f}_n^{\rm ext+}$ are the discrete generalized external forces evaluated as
\begin{align}
\mathbf{f}_n^{\rm ext-}=\frac{\Delta t}{2} \mathbf{f}^{\rm vis} (\frac{\mathbf{q}_{n+1}+\mathbf{q}_n}{2},\frac{\mathbf{q}_{n+1}-\mathbf{q}_n}{\Delta t}),\qquad
\mathbf{f}_n^{\rm ext+}=\frac{\Delta t}{2} \mathbf{f}^{\rm vis} (\frac{\mathbf{q}_{n-1}+\mathbf{q}_n}{2},\frac{\mathbf{q}_{n-1}-\mathbf{q}_n}{\Delta t}).
\end{align}

\subsection{Null space matrix and parametrization of rotations}
The internal null space matrix $\mathbf{P}_{\rm int}$ can be found by expressing the redundant velocity $\dot{\mathbf{q}} \in \mathbb{R}^{15}$ in terms of the generalised velocity $\mathbf{t} \in \mathbb{R}^{9}$
\begin{align}
\dot{\mathbf{q}}=\mathbf{P}_{\rm int}(\mathbf{q})\cdot \mathbf{t},
\end{align}
where the generalized velocity is composed of the translational velocity $\dot{\boldsymbol{\varphi}}$, the angular velocity $\boldsymbol{\omega}$ and the velocity of electric potential $\dot{\boldsymbol{\phi}} $, i.e.
$\mathbf{t}=\begin{bmatrix}
\dot{\boldsymbol{\varphi}} & \boldsymbol{\omega} & \dot{\boldsymbol{\phi}} 
\end{bmatrix}^T$. The corresponding internal null space matrix at time $t_n$ is written as
\begin{align}
\mathbf{P}_{\rm int}(\mathbf{q}_n)=
\begin{bmatrix}                           
\mathbf{I} & \mathbf{0} & \mathbf{0}\\                                               
\mathbf{0} & -\hat{\mathbf{d}}_{1,n} & \mathbf{0}\\   
\mathbf{0} & -\hat{\mathbf{d}}_{2,n} & \mathbf{0}\\  
\mathbf{0} & -\hat{\mathbf{d}}_{3,n} & \mathbf{0}\\
\mathbf{0}& \mathbf{0}& \mathbf{I}
\end{bmatrix}, \label{null}
\end{align}
where $\hat{\mathbf{d}}_{i,n}$ denotes the skew-symmetric matrix corresponding to the director vector $\mathbf{d}_{i,n}$ {at $t_n$} and $\mathbf{I}$ is the 3 by 3 identity matrix. {For a multibody dynamic system composed of flexible beam actuators, rigid bodies, joints and constraints, the null space matrix can be designed by considering the electric potential as extra degree of freedom as well.}

To solve the {discrete Euler–Lagrange equations} efficiently, the system can be reduced further into the minimal possible dimensions by use of the nodal reparametrisation, which can be achieved by introducing a rotation matrix $\mathbf{R}(\boldsymbol\theta)$ parametrised in terms of the rotational variable $\boldsymbol\theta$. The rotation matrix can be chosen as the exponential map, see e.g. \citep{marsden2013introduction},
\begin{align}
\mathbf{R}(\boldsymbol\theta)={\rm exp}(\hat{\boldsymbol\theta})=\mathbf{I}+\frac{{\rm sin} \left\| \boldsymbol\theta \right\| }{\left\| \boldsymbol\theta \right\| } \hat{\boldsymbol\theta} + \frac{1}{2} \left( \frac{{\rm sin} \left\| \boldsymbol\theta \right\| /2}{\left\| \boldsymbol\theta \right\| /2}  \right)^2 (\hat{\boldsymbol\theta})^2.
\end{align}
The generalized configuration of the electromechanically coupled beam is specified by
\begin{align}
\mathbf{u}=\begin{bmatrix} \mathbf{u}_\varphi& \boldsymbol\theta&\mathbf{v}\end{bmatrix}^T
\end{align}
with $\mathbf{u}_\varphi,  \boldsymbol\theta$ and $\mathbf{v}$ characterizing the incremental displacement, the incremental rotation and the incremental electric potential, respectively. In this case, the nodal configuration for the next time step can be updated as
\begin{align}
\mathbf{q}_{n+1} = \mathbf{F}_d ({\mathbf{u}_{n+1}}, \mathbf{q}_{n})=
\begin{bmatrix}
\boldsymbol{\varphi}_n+\mathbf{u}_\varphi\\
{\rm exp}(\hat{\boldsymbol\theta})\cdot \mathbf{d}_{1,n}\\
{\rm exp}(\hat{\boldsymbol\theta})\cdot \mathbf{d}_{2,n}\\
{\rm exp}(\hat{\boldsymbol\theta})\cdot \mathbf{d}_{3,n}\\
\boldsymbol{\phi} + \mathbf{v}
\end{bmatrix}. \label{update}
\end{align}

\subsection{Tangent matrix}
By means of the nodal reparametrisation, the unknowns of the equations system in Eq. (\ref{R}) is changed from $\mathbf{q}_{n+1}$ to the generalized variables $\mathbf{u}_{n+1}$. The nonlinear equation system is solved by use of the  Newton-Rapson algorithm with the tangent matrix at iteration $i$
\begin{align}
\mathbf{K}_T^i
=\mathbf{P}^T(\mathbf{q}_n)  \frac{\partial \mathbf{R}^L(\mathbf{q}_{n+1}^i)}{\partial \mathbf{q}_{n+1}^i} \frac{\partial \mathbf{q}_{n+1}(\mathbf{u}_{n+1}^i)}{\partial \mathbf{u}_{n+1}^i},
\end{align}
in which $\mathbf{R}^L(\mathbf{q}_{n+1})$ is the residual of the discrete Euler–Lagrange equation. When boundary conditions are imposed on some degrees of freedom of a beam node, the corresponding components in null space matrix will be zero. By crossing out the rows involved in boundary conditions, the non-singular tangent matrix can be obtained. Accordingly, the boundary conditions can be imposed by setting specific incremental values in the update of the nodal configuration in Eq. (\ref{update}). In this work, the residual vector and the tangent matrix are derived by using the automatic differentiation tool CasADi \citep{Andersson2019}.

\subsection{Legendre transformation for energy evaluation and system initialization}
{Since the velocity involved in the kinetic energy is approximated with the finite difference scheme, it does not fulfill the hidden constraints and their time derivatives exactly.} An alternative energy formulation is the discrete Hamiltonian evaluated with the momentum, which can be obtained via the discrete Legendre transformation, see e.g. \citep{leyendecker2008variational}. Since the consistent mass matrix corresponding to the kinetic energy in Eq. (\ref{T}) is singular, it can not be applied to compute the Hamiltonian directly where the inverse of {the} mass matrix is required. Due to the fact that the kinetic energy is independent of the rate of $\mathbf{d}_3$ and the rate of electric potential, {a} reduced non-singular mass matrix $\bar{\mathbf{M}}$ can be {defined at node $s$ as}
\begin{align}
\bar{\mathbf{M}}=\begin{bmatrix}                           
A_\rho\mathbf{I} & \mathbf{0} & \mathbf{0}\\                                               
\mathbf{0} & M_\rho^1\mathbf{I} & \mathbf{0}\\   
\mathbf{0} & \mathbf{0} & M_\rho^2\mathbf{I}
\end{bmatrix}
\end{align}
with $\mathbf{I}$ the $3 \times 3$ identity matrix. Therefore, the discrete Hamiltonian can be evaluated with the reduced mass matrix $\bar{\mathbf{M}}$ and its conjugated reduced momentum
\begin{align}
\bar{\mathbf{p}}=[\mathbf{p}_{\varphi}, \mathbf{p}_1, \mathbf{p}_2 ]^T.
\end{align}
To eliminate the Lagrangian multiplier term from the reduced momentum $\bar{\mathbf{p}}$, the projected discrete Legendre transformation can be applied. In this case, the projected reduced momentum can be computed as
\begin{align}
^Q\bar{\mathbf{p}}_n^-=-\bar{\mathbf{Q}}(\mathbf{q}_{n})\frac{\partial L_d(\mathbf{q}_{n},\mathbf{q}_{n+1})}{\partial \bar{\mathbf{q}}_{n}},
\end{align}
where $^Q\bar{\mathbf{p}}_n^-$ is the reduced momentum at time $t_n$, $\bar{\mathbf{Q}}$ is the projection matrix and {$\bar{\mathbf{q}}_{n}$ is the vector containing the first 9 elements of $\mathbf{q}_{n}$}. Corresponding to the reduced momentum $^Q\bar{\mathbf{p}}_n^-$ and the mass matrix $\bar{\mathbf{M}}$, the projection matrix is computed in the reduced form
\begin{align}
\bar{\mathbf{Q}}=\mathbf{I}_{9\times 9}-\bar{\mathbf{G}}^T(\bar{\mathbf{G}}\bar{\mathbf{M}}^{-1}\bar{\mathbf{G}}^T)^{-1}\bar{\mathbf{G}}\bar{\mathbf{M}}^{-1}
\end{align}
with the reduced internal constraint Jacobi matrix $\bar{\mathbf{G}}_{3 \times 9}$ given by
\begin{align}
\bar{\mathbf{G}}
=\begin{bmatrix}                           
\mathbf{0} &   \mathbf{d}_1^T&  \mathbf{0} \\
\mathbf{0} &   \mathbf{0} &  \mathbf{d}_2^T \\
\mathbf{0} &   \mathbf{d}_2^T &  \mathbf{d}_1^T \\
\end{bmatrix}.
\end{align}
The discrete Hamiltonian $H_d$ can be computed as 
\begin{align}
H_d =\int_c \left( \frac{1}{2} {^Q\bar{\mathbf{p}}_n^{-T}} \cdot \bar{\mathbf{M}}^{-1} \cdot {^Q\bar{\mathbf{p}}_n^-}\right) ds + V(\mathbf{q}_n).
\end{align}

Furthermore, the discrete Legendre transformation is applied to initialize the system at time step $t_0$. In the discrete Euler–Lagrange equation (\ref{R}), the unknown value of $ \mathbf{q}_{-1}$  is required to solve $ \mathbf{q}_{1}$. To avoid the computation of $ \mathbf{q}_{-1}$, the discrete momentum obtained from the Legendre transformation is initialized with an initial momentum $\mathbf{p}_{0}^-=\mathbf{p}(0)$, which leads to the equation of motion at $t_0$
\begin{align}
\mathbf{P}^T(\mathbf{q}_0) \left[ -\frac{\partial L_d(\mathbf{q}_{0},\mathbf{q}_{1})}{\partial \mathbf{q}_{0}} -\mathbf{f}_0^{\rm ext-}\right] =\mathbf{P}^T(\mathbf{q}_0) \mathbf{p}(0).
\end{align}
In this case, $\mathbf{q}_{1}$ can be {computed given $ \mathbf{q}_{0}= \mathbf{q}(0)$ and $\mathbf{p}(0)$.}

\section{Numerical examples}
The material parameters of the dielectric elastomer applied in this work are shown in Table 1. Instead of using the Lam{\'e} parameter $\mu=0.233Mpa$ in \citep{schlogl2016electrostatic} for the nearly incompressible case, the Lam{\'e} parameter $\mu$ is set to be $233Mpa$ in this work for the sake of a smaller Poisson's ratio. According to the discussions in Section 5.1, the smaller Poisson's ratio allows {the better comparison of 3D FEM model with the beam model with assumption of no lateral contractions.}

\begin{table}[htb!]
	\centering
	\caption{Material parameters} 
	\begin{tabular}{ccccccc}
		\hline
		$\rho$& $\lambda$ &$\mu$ &$\varepsilon_{0}$ & $c_1$ & $c_2$  \\
		\hline
		$g/mm^3$ & $Mpa$ & $Mpa$ & $C/Vm$ & $N/V^2$ & $N/V^2$\\
		\hline
		$1$ &  $999.8$  & $233$ & $8.854\times 10^{-12}$ &  $5\times 10^{-8}$  &  $1\times 10^{-9}$ \\
		\hline
	\end{tabular}
\end{table}

\begin{figure}[htb!]
	\centering
	\includegraphics[width=.35\textwidth]{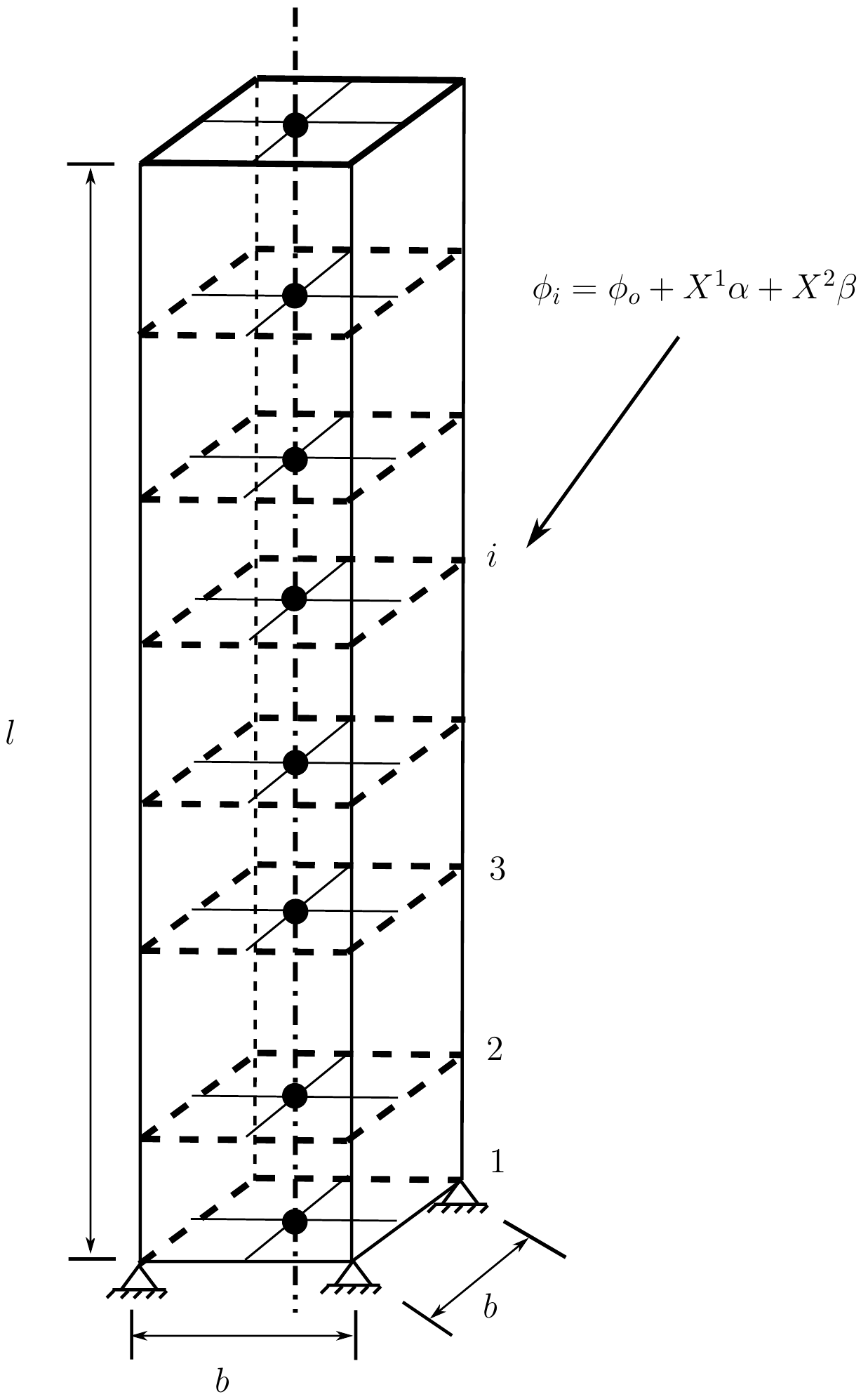}
	\caption{Boundary conditions on the beam.}
	\label{beam_geo}
\end{figure}

The geometry of the beam with length $l$ and square cross section $b \times b$ is shown in Fig. \ref{beam_geo}. The beam is fixed at one end. The electrical boundary condition $\phi_i$ is imposed on node $i$ such that different deformations on beam can be generated, including uniaxial contraction, shear, bending and torsion. The electrical boundary conditions $\phi_i$ and the beam size for generating different beam deformations in this work are given in Table 2.

\begin{table}[htb!]
	\centering
	\caption{electrical boundary conditions and beam size} 
	\begin{tabular}{llllllllll}
		\hline
		& $\phi_i/V$ & $\phi_o$ & $\alpha$& $\beta$ & $N_e$&$b/mm$&$l/mm$\\
		\hline
		Contraction & $\phi_o $ at $i=6$ & $2e4$ &$0$ & $0$ & $5$&0.02&0.1\\
		\hline
		Bending & $0$ at $ i=1,3...;$ $\phi_o+\alpha X^1+\beta X^2$ at $ i=2,4,...$ & $2e2$ &$2e4$ & $0$ & $40$&0.005&0.1\\
		\hline
		Shear & $i\phi_o+\alpha X^1+\beta X^2$ & $2e2$ & $2e4$  & $0$ & $40$&0.005&0.1\\
		\hline
		Torsion & $\phi_o+\alpha(cos\theta_i-sin\theta_i)X^1 + \beta(cos\theta_i+sin\theta_i)X^2$ & $2e3$ & $3e4$  & $3e4$ & $80$&0.005&0.05\\
		\hline
	\end{tabular}
\end{table}

\subsection{Uniaxial contraction}
To generate the uniaxial contraction in the beam, the uniform electric potential is applied on the cross section, i.e. $\alpha, \beta=0$. As shown in Table 2, electric potential $\phi_1=0 V$ is applied to the node at the fixed end and electric potential $\phi_6=2 \times 10^4 V$ is applied to the top beam node. The motion of the beam discretized with {$N_e=5$ elements, i.e. 6 nodes,} is computed with the time step of $1 \times 10^{-4} ms$.

\begin{figure}[htb!]
	\centering
	\subfigure[]{\label{4a}\includegraphics[width=.44\textwidth]{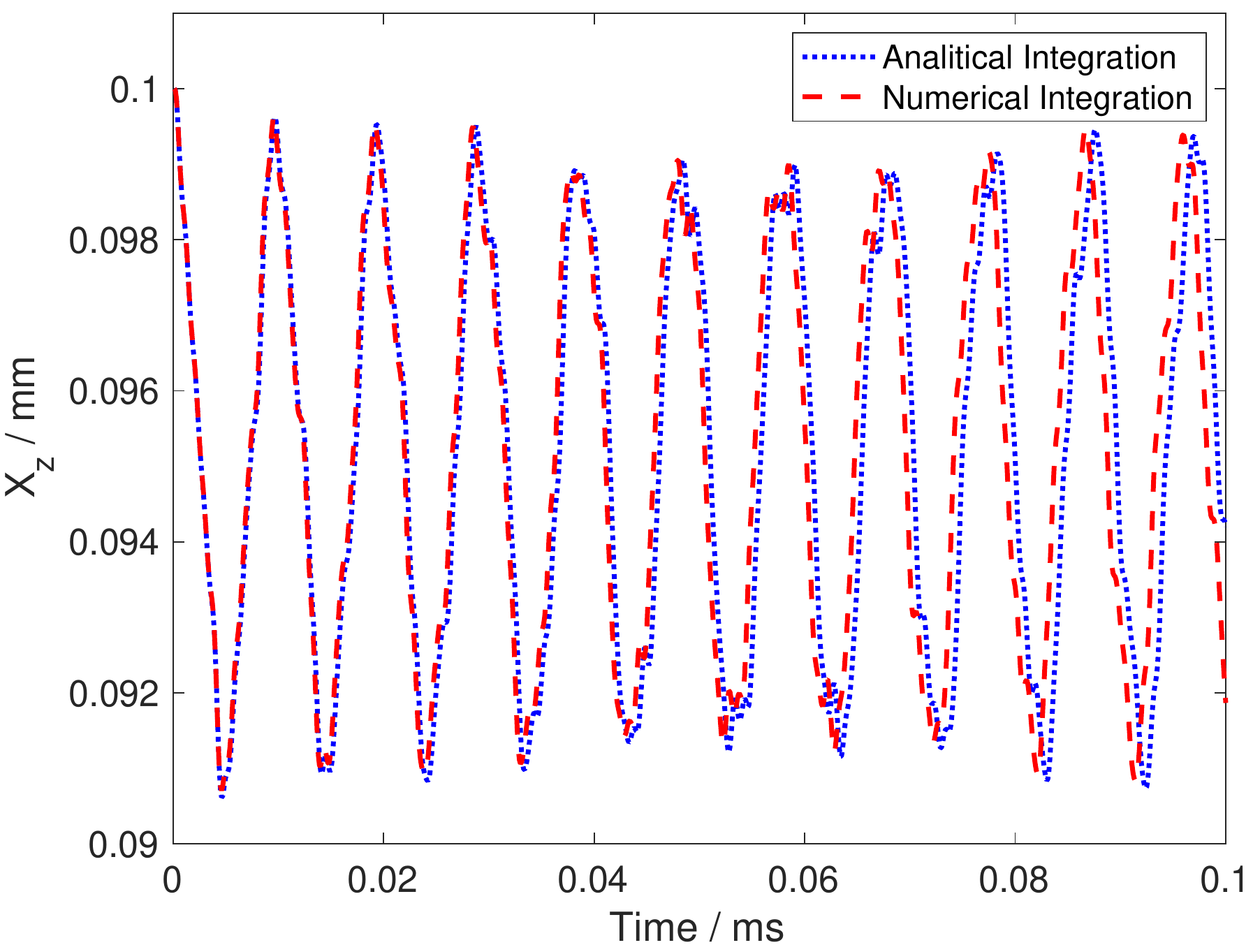}}
	\quad
	\subfigure[]{\label{4b}\includegraphics[width=.45\textwidth]{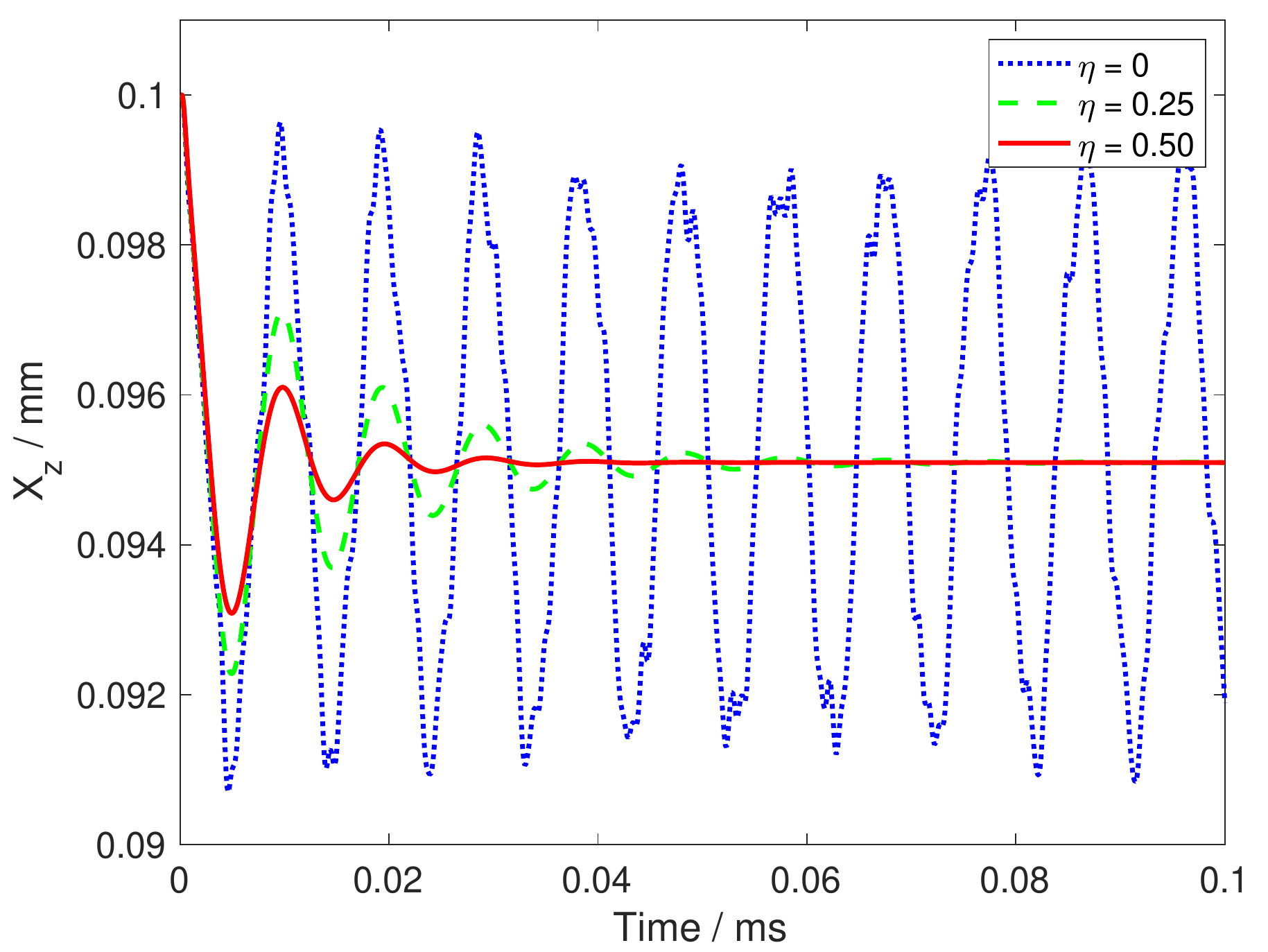}}
	\caption{(a) {Validation of the derived strain energy} and (b) the damping effect on the top beam node.}
	\label{end_node1}
\end{figure}

{Due to the contractive forces from the dielectric effect, the vibration of the beam is induced as shown in Fig. \ref{end_node1}, where the position $X_z$ of the beam node at the free end is depicted. In Fig. \ref{4a}, the beam strain energy $\Omega_b$ derived in the Appendix is compared with that evaluated by the numerical integration over cross section. The derived strain energy shows good agreement with the numerical model in Fig. \ref{4a} and will be applied in the following parts.} In Fig. \ref{4b}, the damping effect with different values of the viscosity parameter $\eta$ is depicted. Without viscosity i.e. $\eta=0$, the vibration of the beam continues over time. With the increase of viscoelastic effect, the increase of damping effect on the beam can be observed. By setting $\eta=0.5$, the viscoelastic beam is gradually approaching a static state.\\

\begin{figure}[htb!]
	\centering
	\subfigure[Discrete Hamiltonian $H_d$]{\label{5a}\includegraphics[width=.45\textwidth]{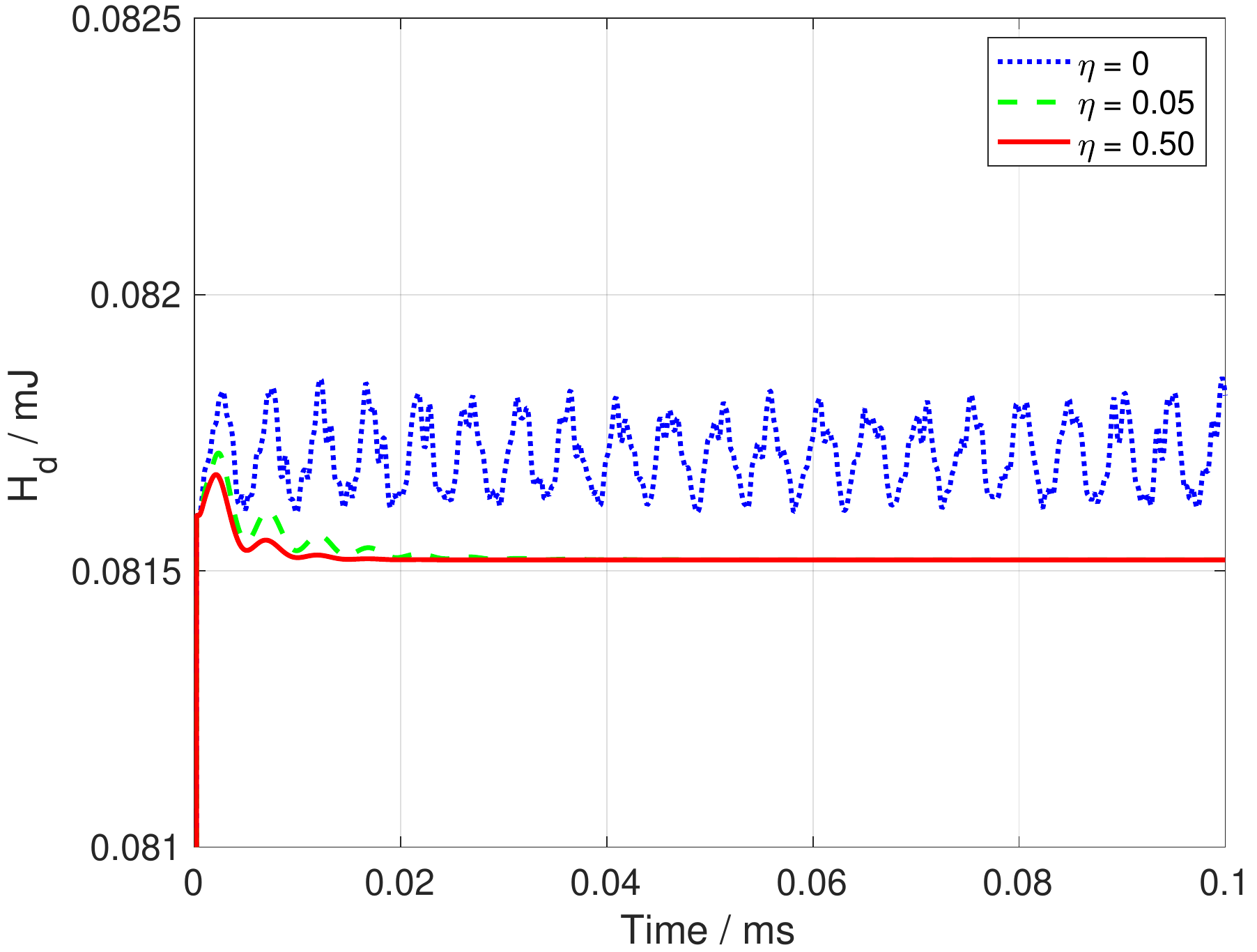}}
	\hspace{0.5cm}
	\subfigure[Discrete potential energy and kinetic energy in $H_d$]{\label{5b}\includegraphics[width=.48\textwidth]{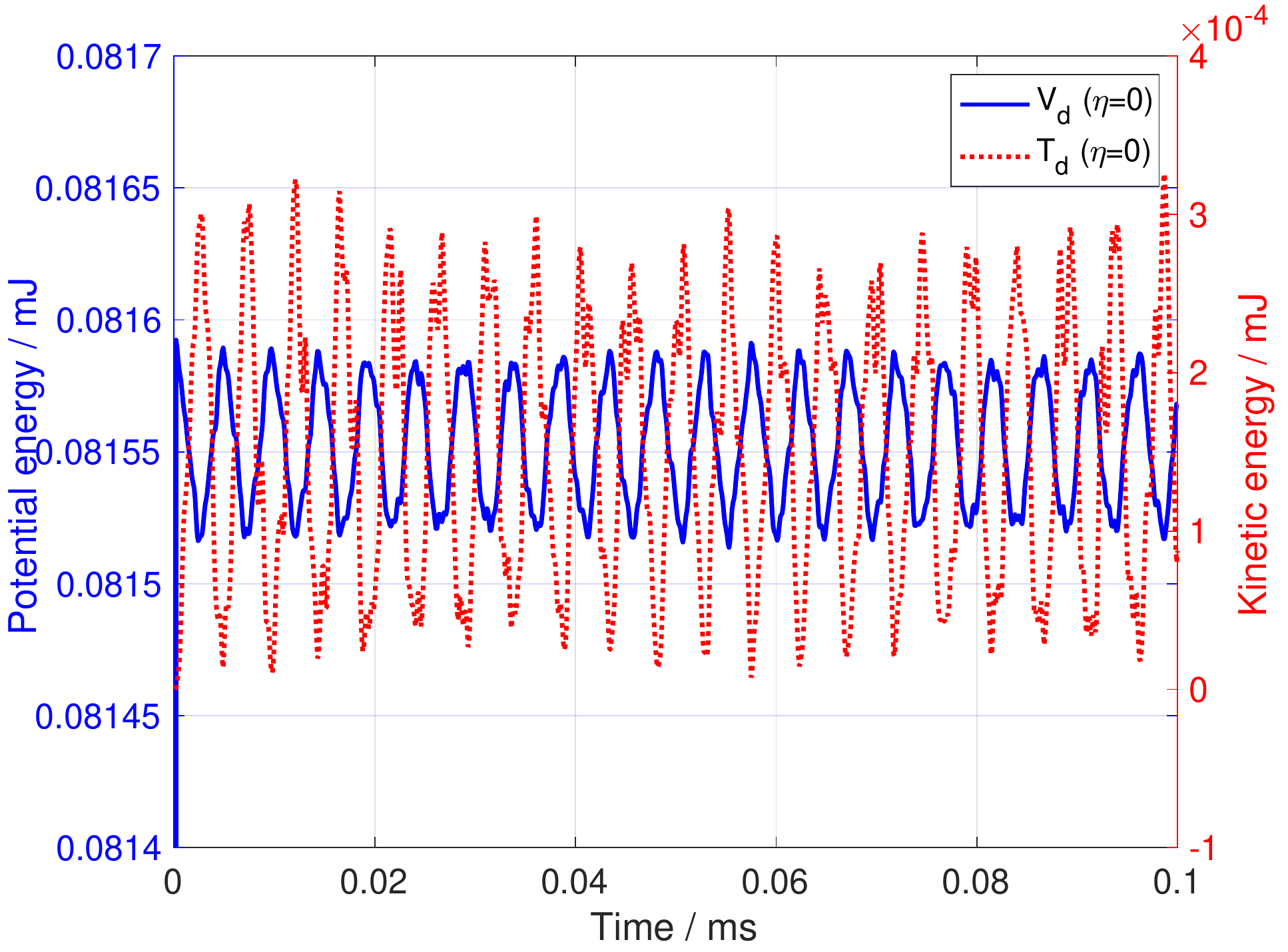}}\\
	\caption{Energy behavior of the beam.}
	\label{EH}
\end{figure}

To evaluate the energy property of the beam, the discrete Hamiltonian $H_d$ computed with momentum is plotted in Fig. \ref{EH}. As shown in Fig. \ref{5a}, the total energy oscillates around a value when there is no viscosity i.e. $\eta=0$. By introducing the viscosity, the total energy decreases at the beginning stage and then preserves a smaller value, where the decrease of total energy is induced by the decrease of kinetic energy. Additionally, with the increase of the damping parameter $\eta$, the total energy decreases faster. To evaluate the energy behavior for $\eta=0$ further, the kinetic energy $T_d$ and the potential energy $V_d$ corresponding to the total energy $H_d$ are plotted in Fig. \ref{5b}. It can be observed that the frequency of the kinetic energy is twice as much as the frequency of the displacement in Fig. \ref{end_node1}, which is reasonable by considering the velocity changes. However, the potential energy shares the same frequency with the kinetic energy, which is different from the pure elastic case where the elastic potential energy shares the same frequency with displacement. The reason can be attributed to the fact that the potential energy of the dielectric elastomer holding its maximum value in the undeformed state decreases to a minimum value in an intermediate compressed state, and then increases with the further compression, see the first V-shape oscillation of $V_d$ in Fig. \ref{5b}.

\begin{figure}[htb!]
	\centering
	\subfigure{\label{.}\includegraphics[width=\textwidth]{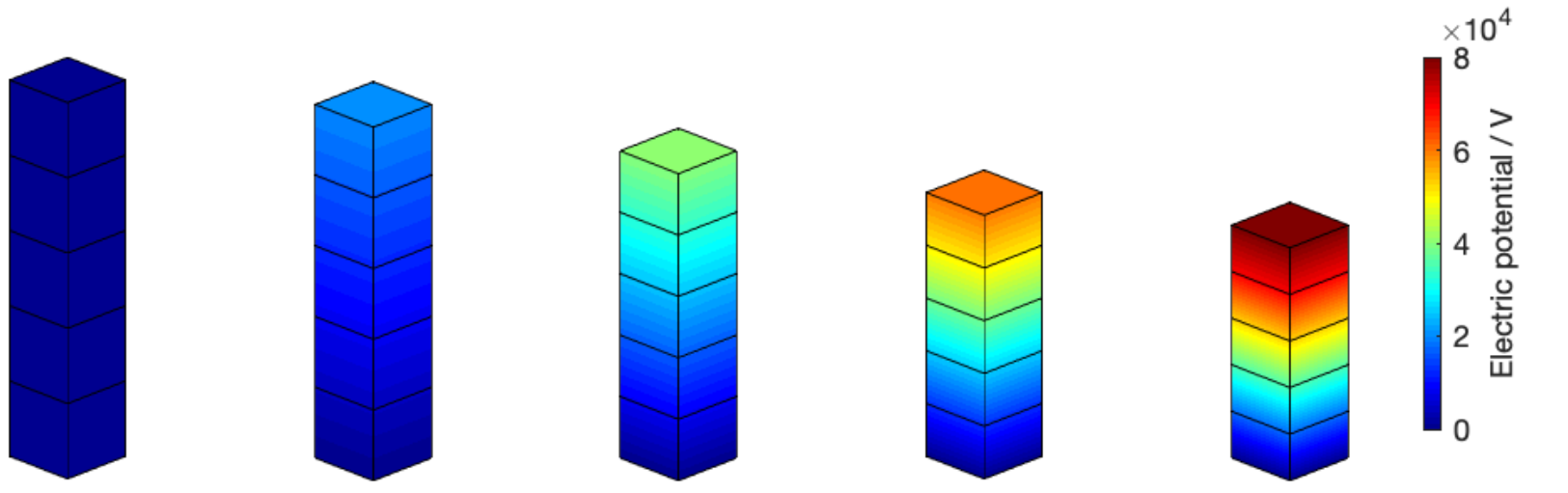}}\\
	\subfigure{\label{.}\includegraphics[width=0.09\textwidth]{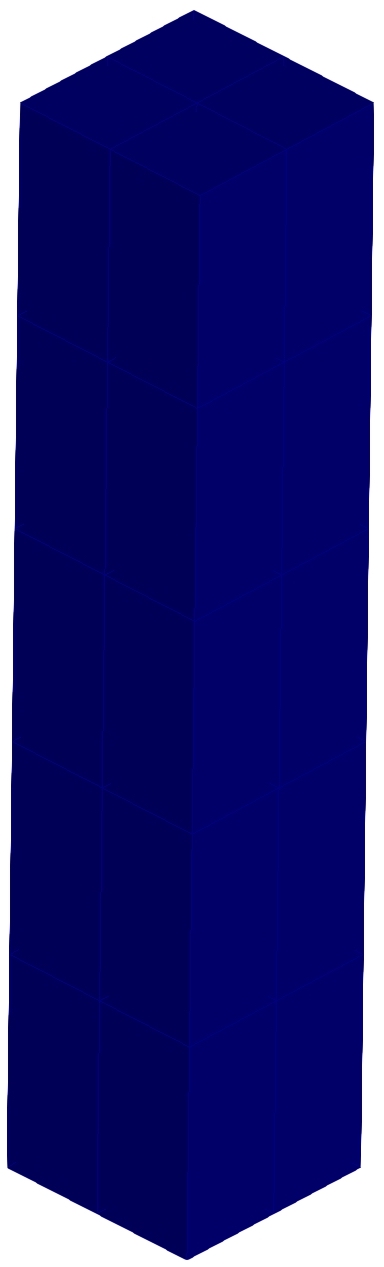}}
	\hspace{1.5cm}
	\subfigure{\label{.}\includegraphics[width=.09\textwidth]{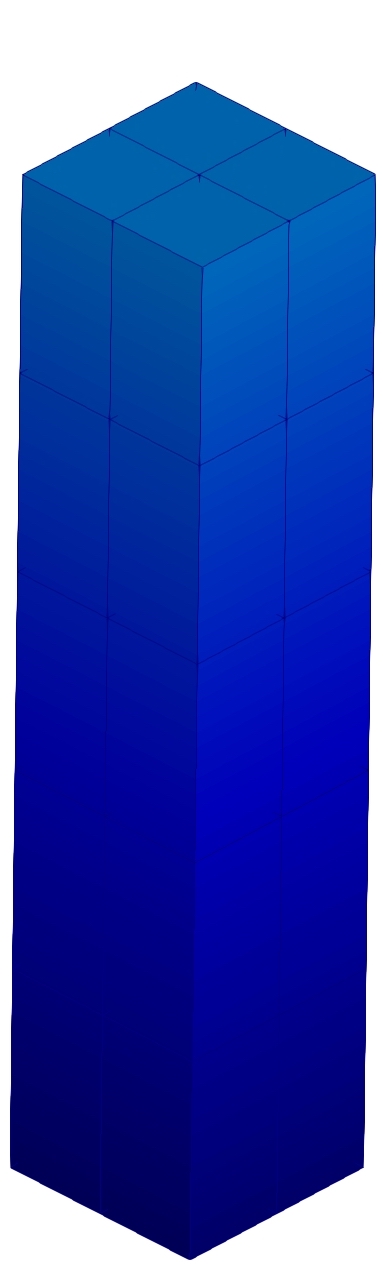}}
	\hspace{1.4cm}
	\subfigure{\label{.}\includegraphics[width=.09\textwidth]{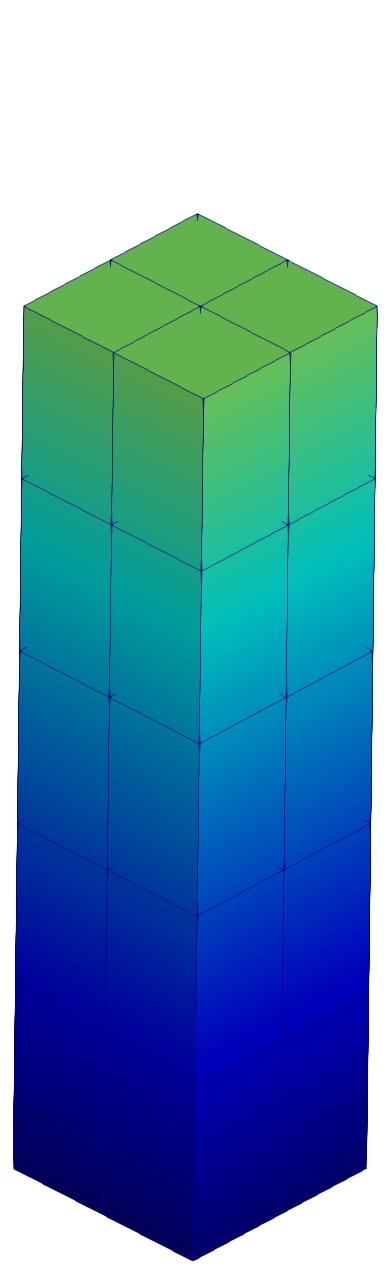}}
	\hspace{1.5cm}
	\subfigure{\label{.}\includegraphics[width=.09\textwidth]{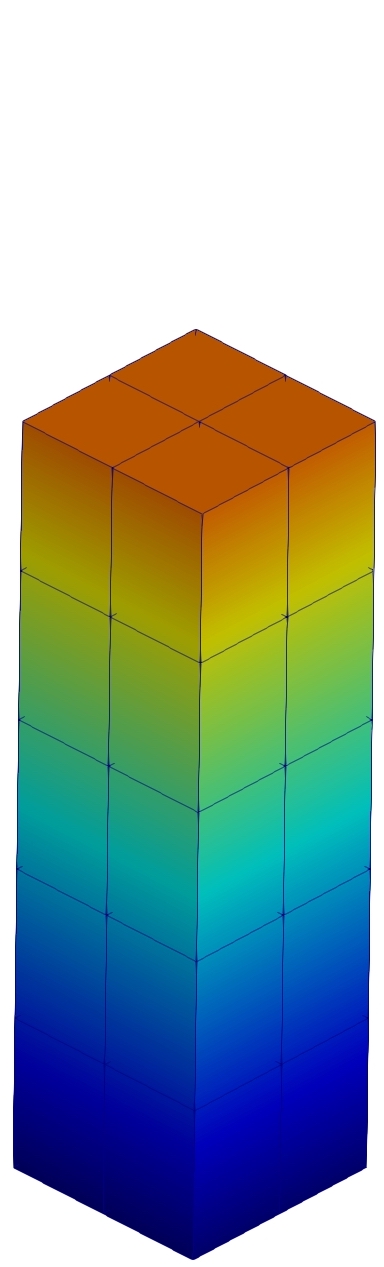}}
	\hspace{1.5cm}
	\subfigure{\label{.}\includegraphics[width=.09\textwidth]{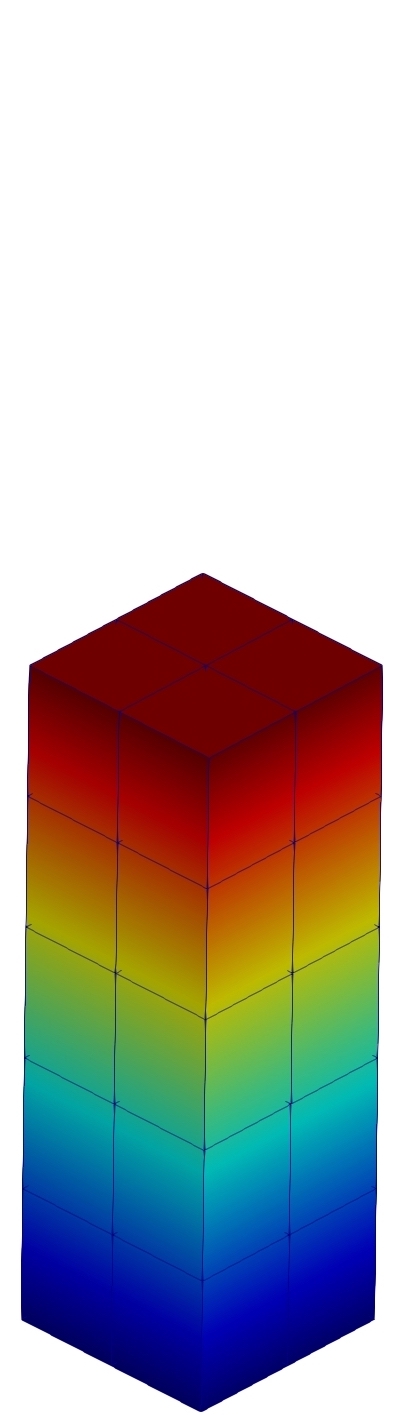}}
	\hspace{.3cm}
	\subfigure{\label{.}\includegraphics[width=.08\textwidth]{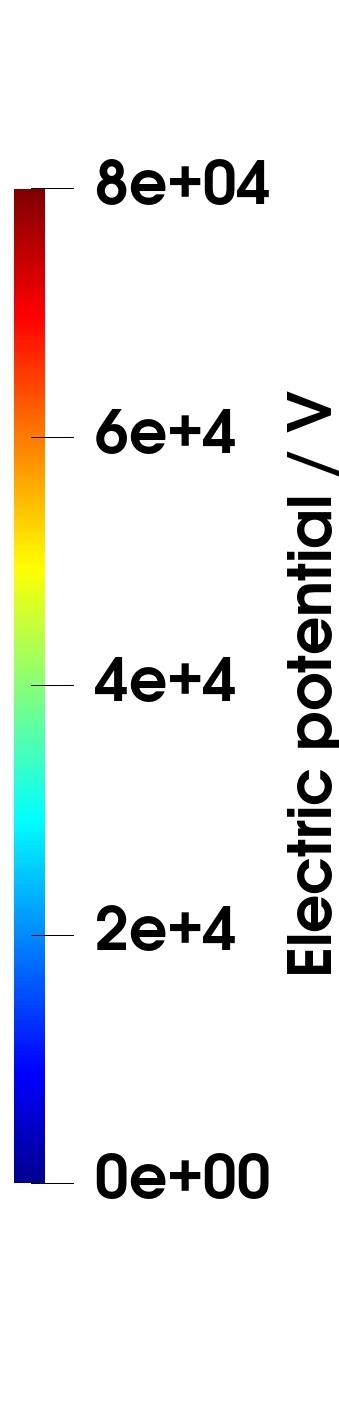}}\\
	\caption{Contractions in the beam model (upper) and the 3D FEM model (lower) with electric potentials $\phi_o=0,2,4,6,8\times 10^4V$.}
	\label{uv}
\end{figure}

The magnitude of contractions of the beam depends on the magnitude of electric potential applied on it. With the increase of electric potential, the increase of contraction can be observed as shown in the upper graph of Fig. \ref{uv}. To validate the beam model, the contractions of the beam model are compared with the results of the 3D FEM model as shown in the lower graph of Fig. \ref{uv}. In the 3D FEM model, the beam structure is discretized with $5$ elements in longitude direction whereas with $2\times2$ elements for the cross section, where the same material model and size as the beam model are applied. To prevent the deformation of cross section, the horizontal movement of the finite element nodes is fixed {to zero}.\\

\begin{figure}[htb!]
	\centering
	\subfigure[]{\label{vu}\includegraphics[width=.49\textwidth]{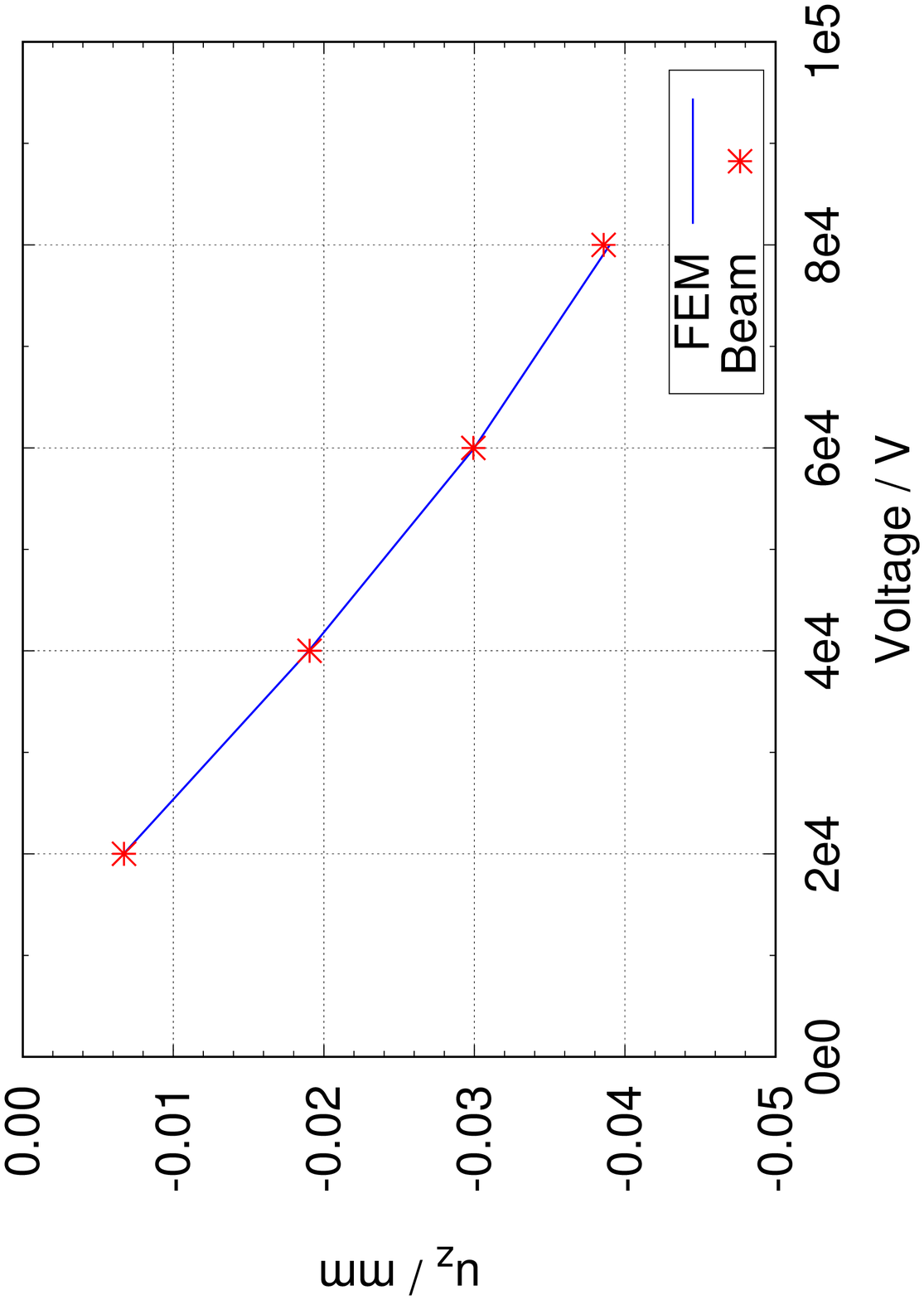}}
	\subfigure[]{\label{dof}\includegraphics[width=.49\textwidth]{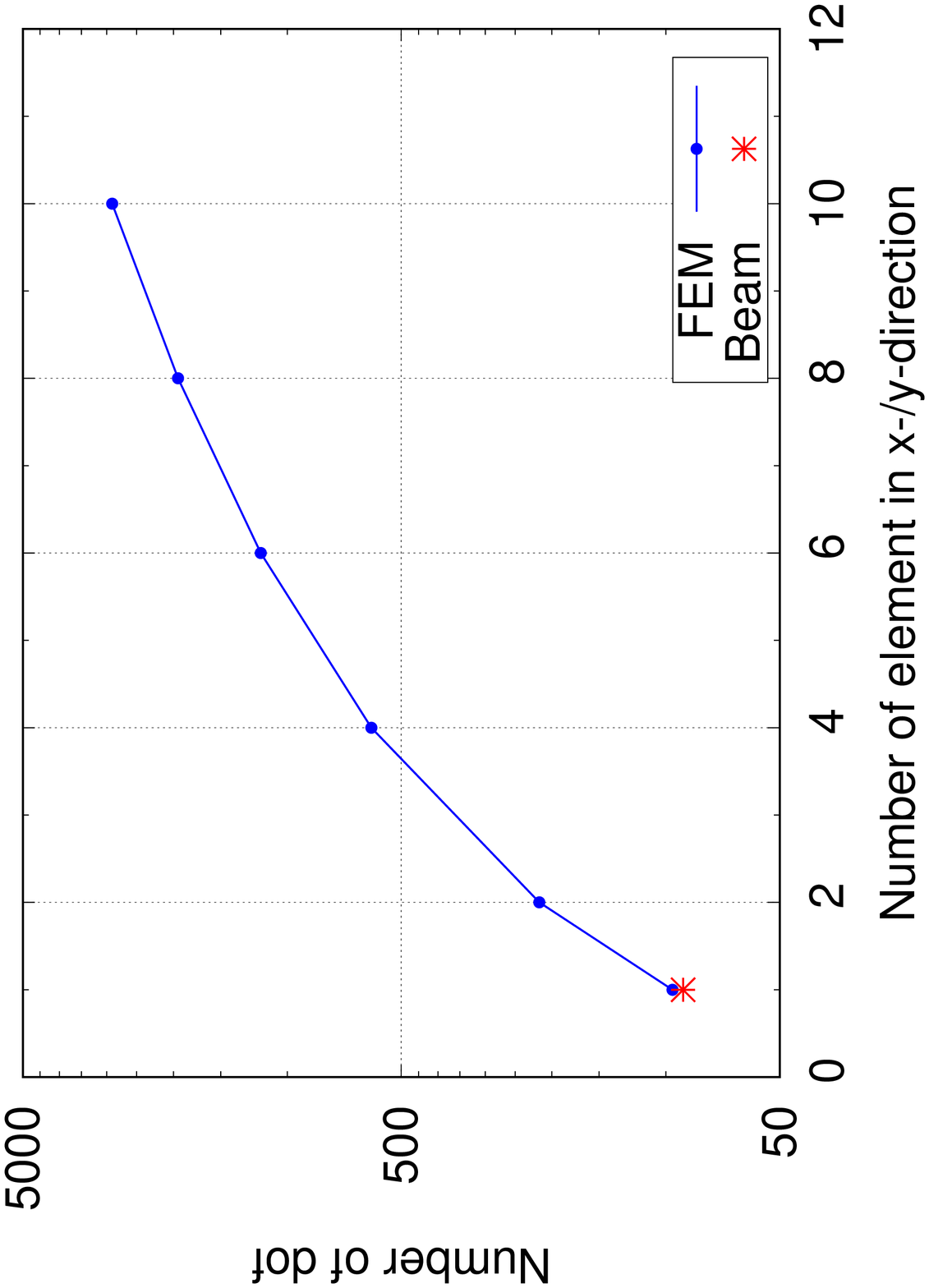}}
	\caption{(a) Displacement of the top beam node in $z$-direction and (b) degree of freedom of the beam.}
\end{figure}

The magnitude of contractions computed by these two approaches is compared in Fig. \ref{vu}. It can be observed that the result of beam model agrees well with the 3D finite element model. In the beam model, 15 degrees of freedom are attached on each node. In the 3D FEM model, {16 degrees of freedom are required at each beam cross section when the cross section is discretized with one element.} However, the number of degree of freedom in the 3D FEM model increases with the increase of number of elements on {the} cross section as shown in Fig. \ref{dof}. It can be observed that, for the same number of elements in the {longitudinal} direction, {less degrees of freedom are required in the beam model.}\\

\subsection{Shear}
\begin{figure}[htb!]
	\centering
	\subfigure[Beam]{\label{currbeam}\includegraphics[width=.24\textwidth]{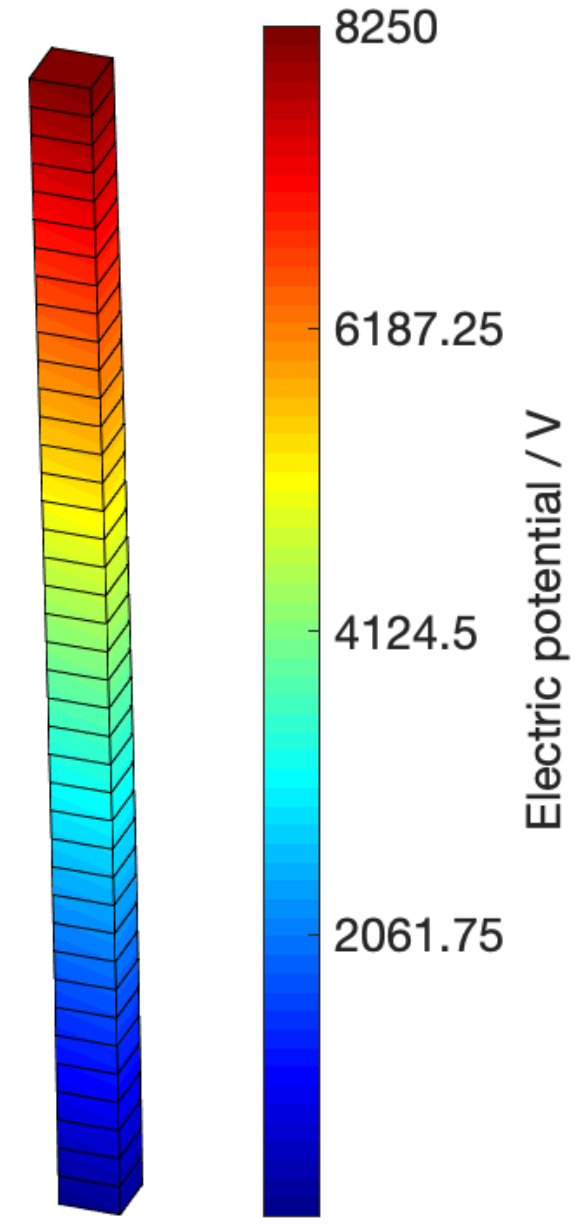}}
	\hspace{2cm}
	\subfigure[3D FEM]{\label{currfem}\includegraphics[width=.205\textwidth]{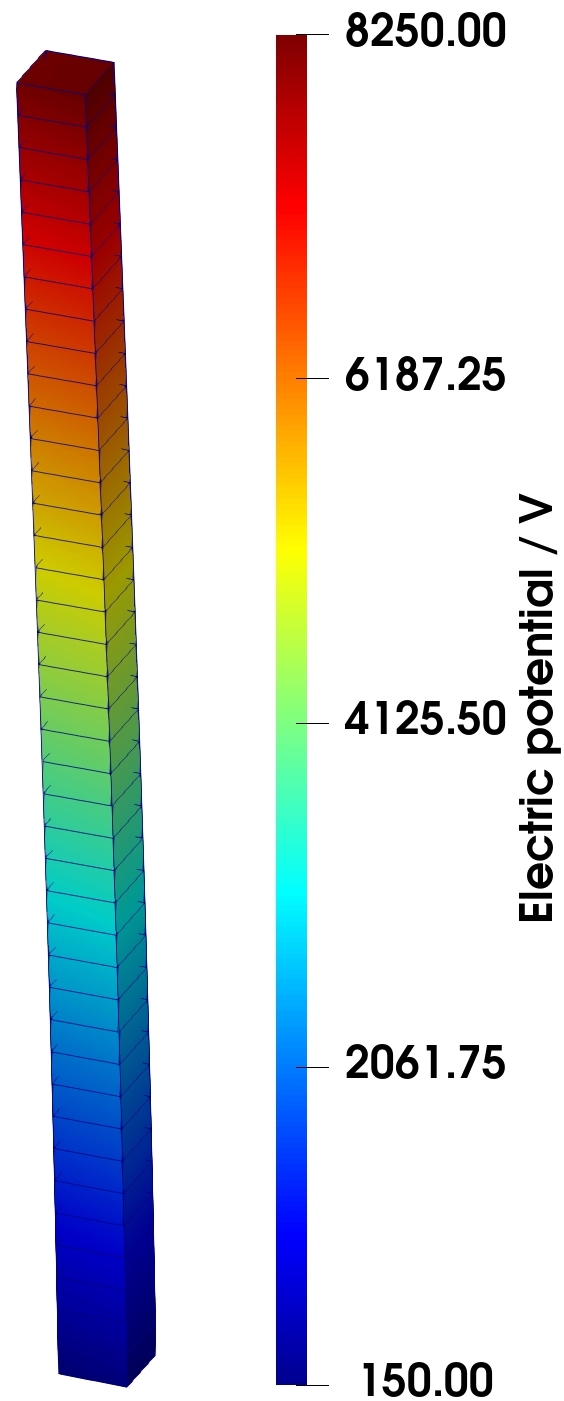}}
	\caption{Shear in the beam model and the 3D FEM model.}
	\label{deform-shear}
\end{figure}

\begin{figure}[htb!]
	\centering
	\includegraphics[width=.4\textwidth]{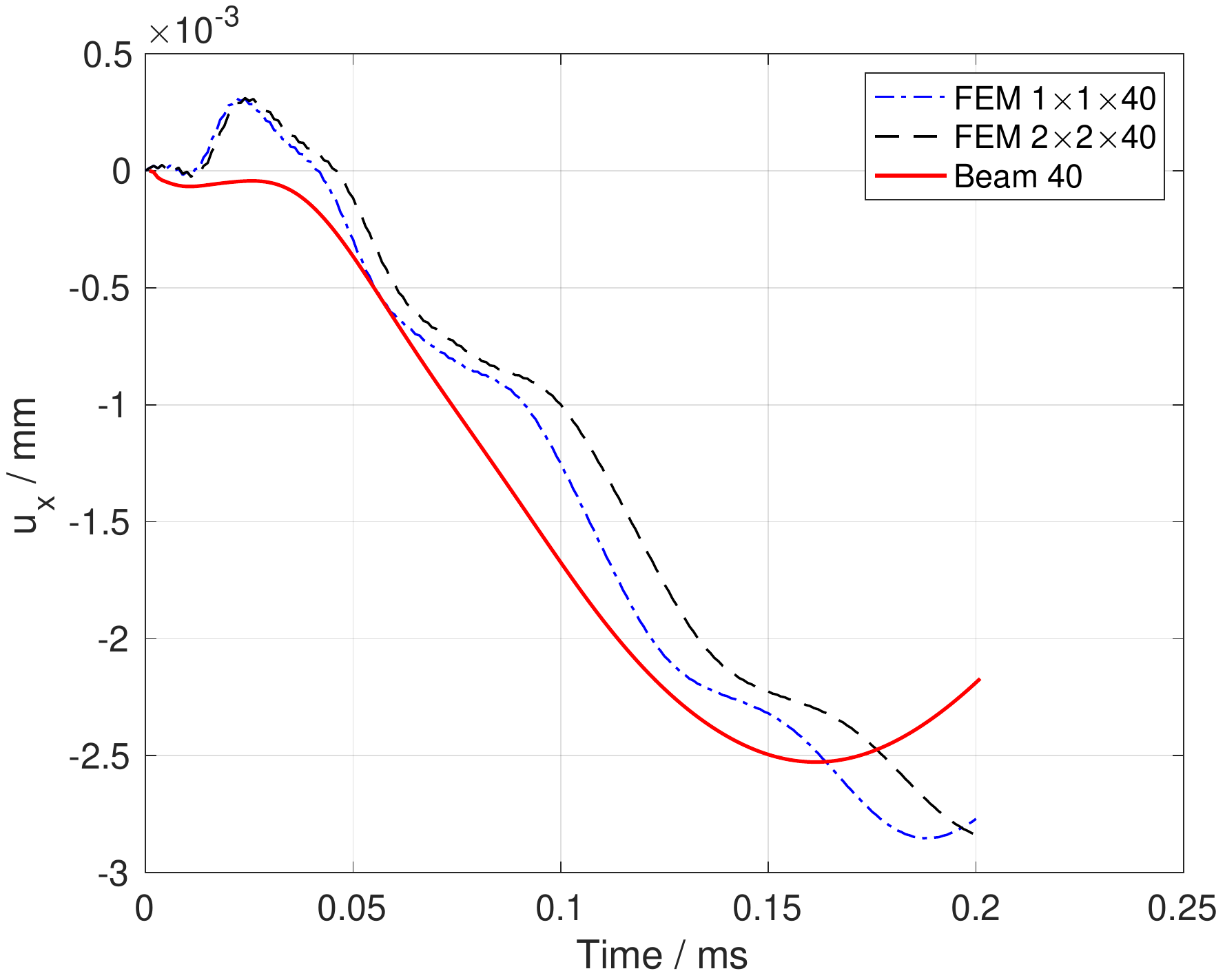}
	\caption{Displacement of the top beam node in $x$-direction.}
	\label{end_nodes}
\end{figure}

Apart form the uniaxial contraction, the electric potential formulated in this work allows more complex beam deformations by applying {a} non-uniform electric potential on the cross sections. To generate shear in the beam, the electric potential $\phi_i=i\phi_o+\alpha X^1+\beta X^2$ is applied to node $i$ as shown in Table 2.  The electric potential on the deformed beam at time $t=0.2ms$ is shown in Fig. \ref{currbeam}, where the beam is discretized with 41 nodes and the damping parameter $\eta$ is set to be $0.5$. By applying the same electrical boundary conditions, the deformation is computed by the 3D FEM model as shown in Fig. \ref{currfem}. The displacement of the node at the free end of the beam is depicted in Fig. \ref{end_nodes}, {where the 3D FEM model is discretized with $1\times1\times40$ and $2\times2\times40$ elements, respectively.} It can be observed that the beam model is close to the 3D FEM model. The deviation can be attributed to the small deformation of the cross section in 3D FEM and the {approximation in deriving strain energy $\Omega_b$, see the Appendix.}\\

\subsection{Bending}
\begin{figure}[htb!]
	\centering
	\subfigure[Beam]{\label{.}\includegraphics[width=.28\textwidth]{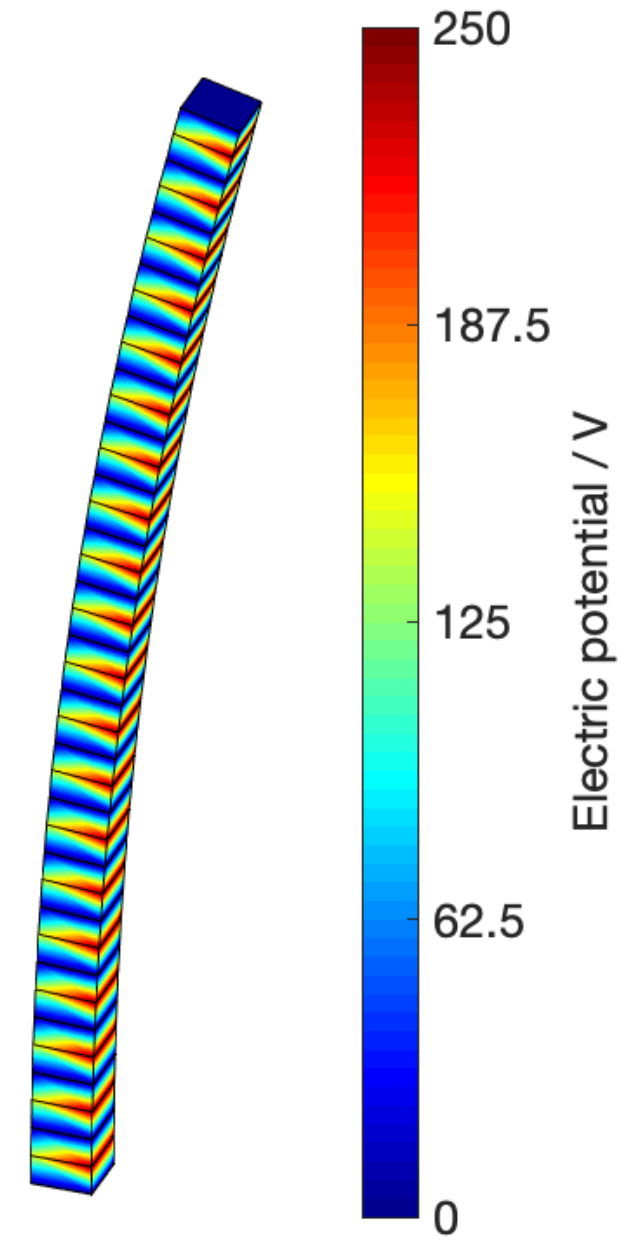}}
	\hspace{2cm}
	\subfigure[3D FEM]{\label{.}\includegraphics[width=.27\textwidth]{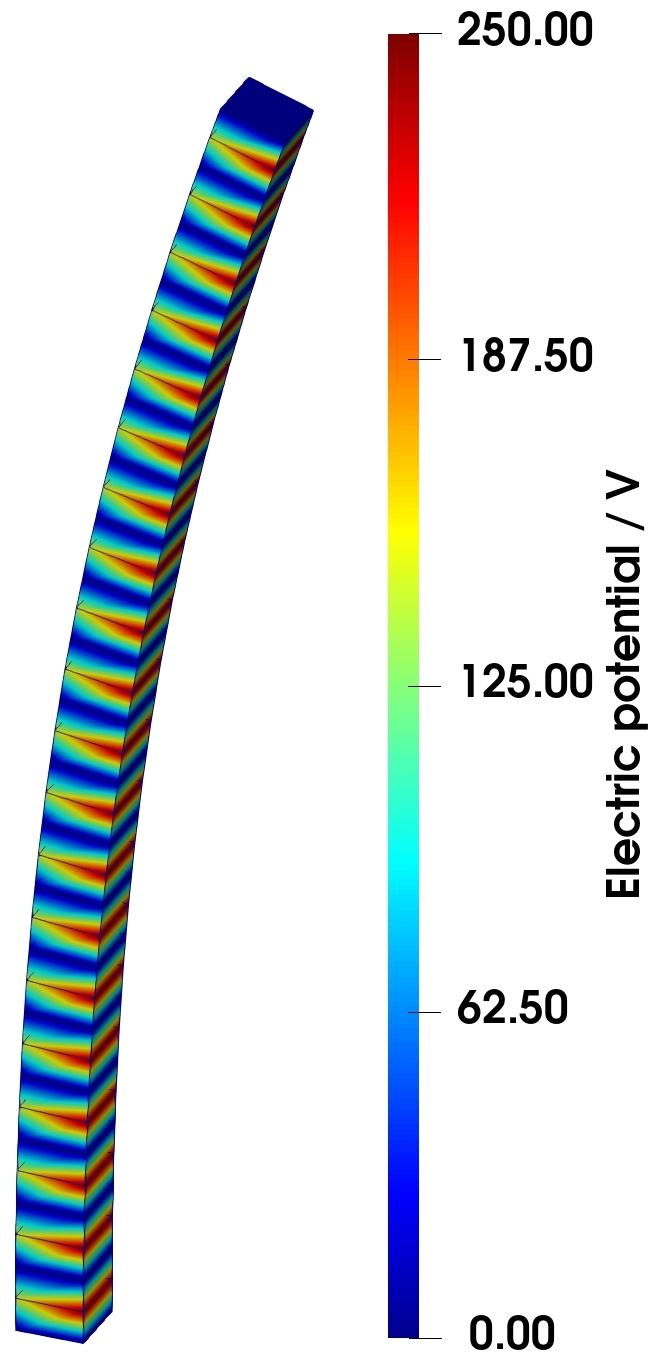}}
	\caption{Bending in the beam model and the 3D FEM model.}
	\label{deform-bending}
\end{figure}

\begin{figure}[htb!]
	\centering
	\includegraphics[width=.4\textwidth]{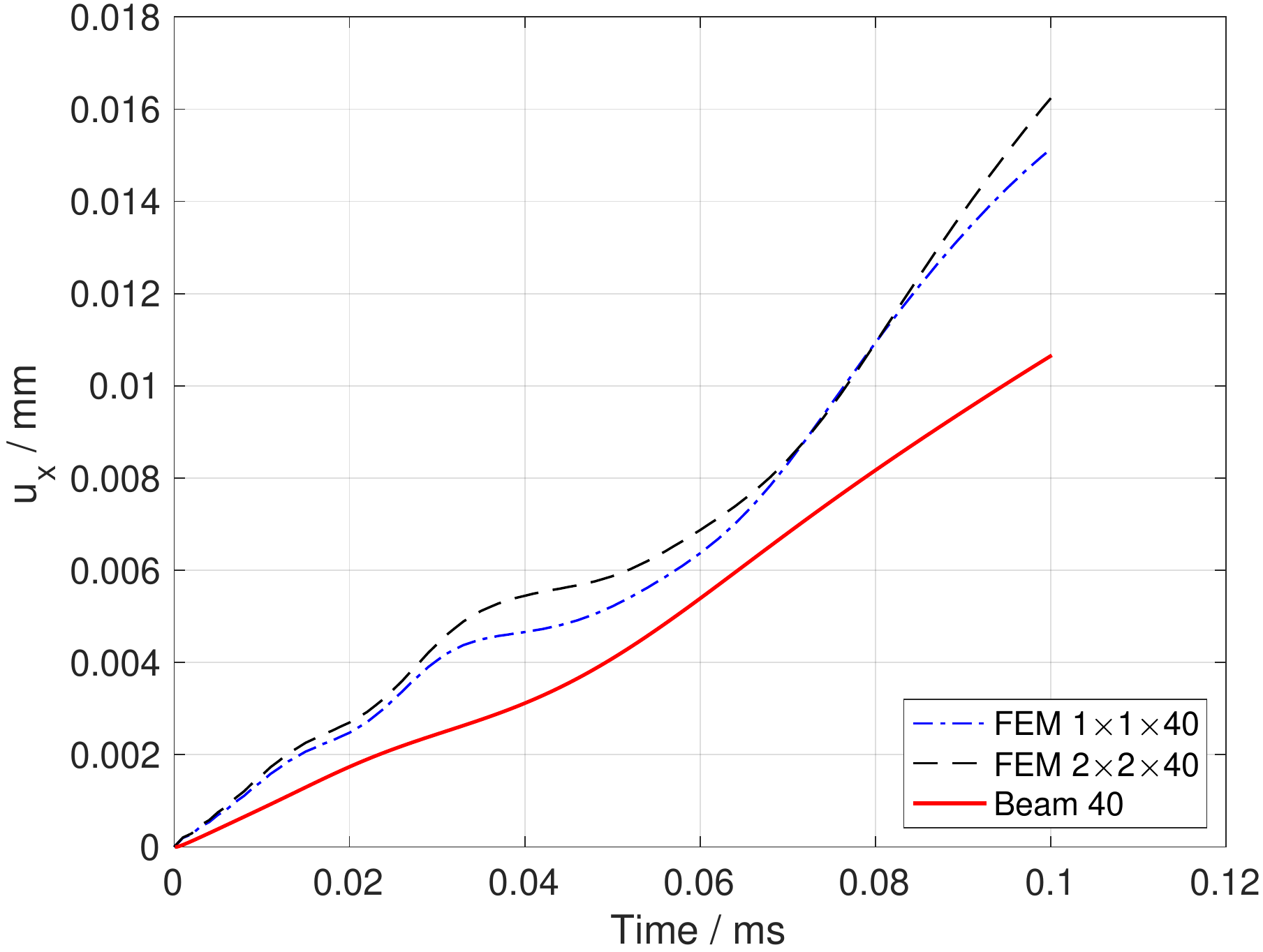}
	\caption{Displacement of the top beam node in $x$-direction.}
	\label{end_node2}
\end{figure}

As shown in Table 2, the bending of the beam can be obtained by alternately applying zero and the non-uniform electric potential $\phi =2 \times 10^2+ 2 \times10^2 X^1$ on the cross sections of beam. In this case, the non-uniform contraction of each beam element leads to the overall bending of beam. The deformed state of the beam model at time $t=0.1ms$ is compared with the 3D FEM model in Fig. \ref{deform-bending}. The displacement of the top beam node {in $x$-direction} is depicted in Fig. \ref{end_node2}, where the agreement between the beam model and the 3D FEM model can be observed.\\

\subsection{Torsion}
\begin{figure}[htb!]
	\centering
	\subfigure[Beam]{\label{.}\includegraphics[width=.216\textwidth]{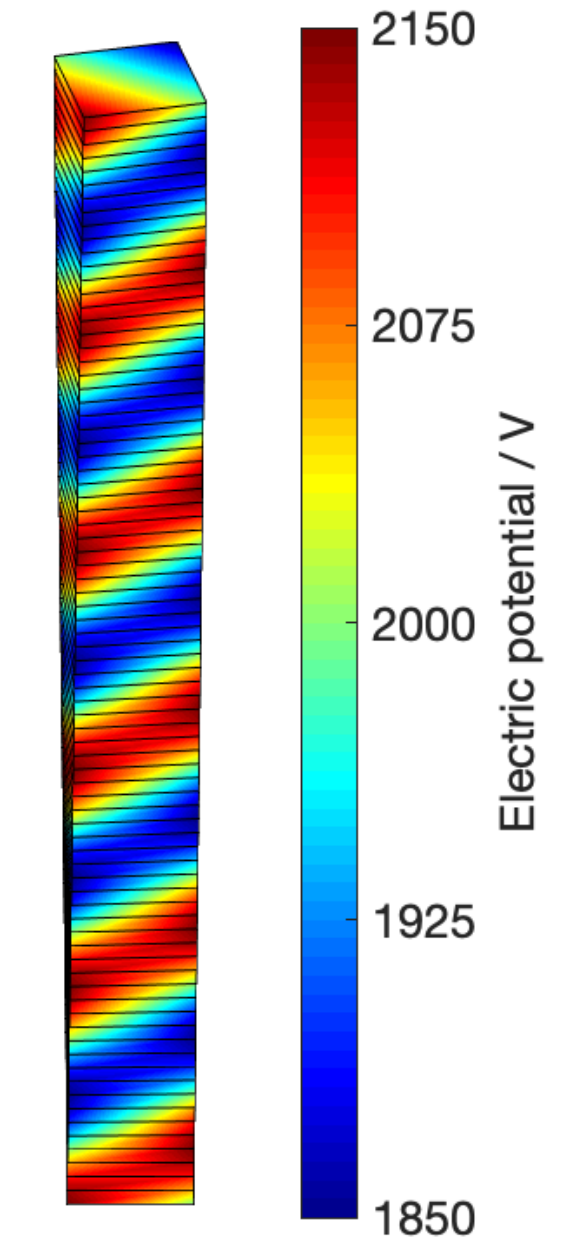}}
	\hspace{2cm}
	\subfigure[3D FEM]{\label{.}\includegraphics[width=.175\textwidth]{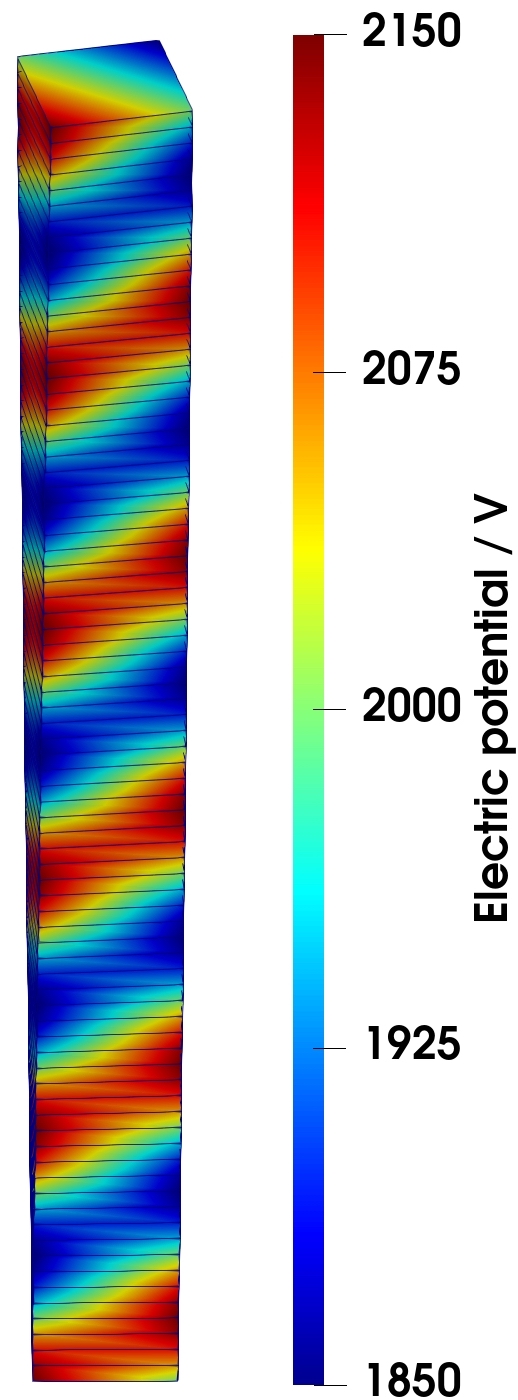}}
	\caption{Torsion in the beam model and the 3D FEM model.}
	\label{deform-torsion}
\end{figure}

The last example is devoted to the torsion {of the} beam where the torque around {the} beam axis has to be generated by applying the electric potential. In this case, the spiral distribution of electric potential can be applied with the electric potential at node $i$ given by
\begin{align}
\phi_i=2 \times 10^3 + 3\times10^4({\rm cos} \theta_i - {\rm sin} \theta_i) X^1 + 3\times10^4({\rm cos} \theta_i + {\rm sin} \theta_i) X^2, \;\;\; \theta_i=(i-1)\frac{\pi}{8}, \;\;\; i=1,2,...
\end{align}
The torsion of the beam model at time $t=0.1ms$ is compared with the 3D FEM model in Fig. \ref{deform-torsion}, where the beam is discretized with 80 elements in the longitudinal direction in both models.\\

\begin{figure}[htb!]
	\centering
	\includegraphics[width=.4\textwidth]{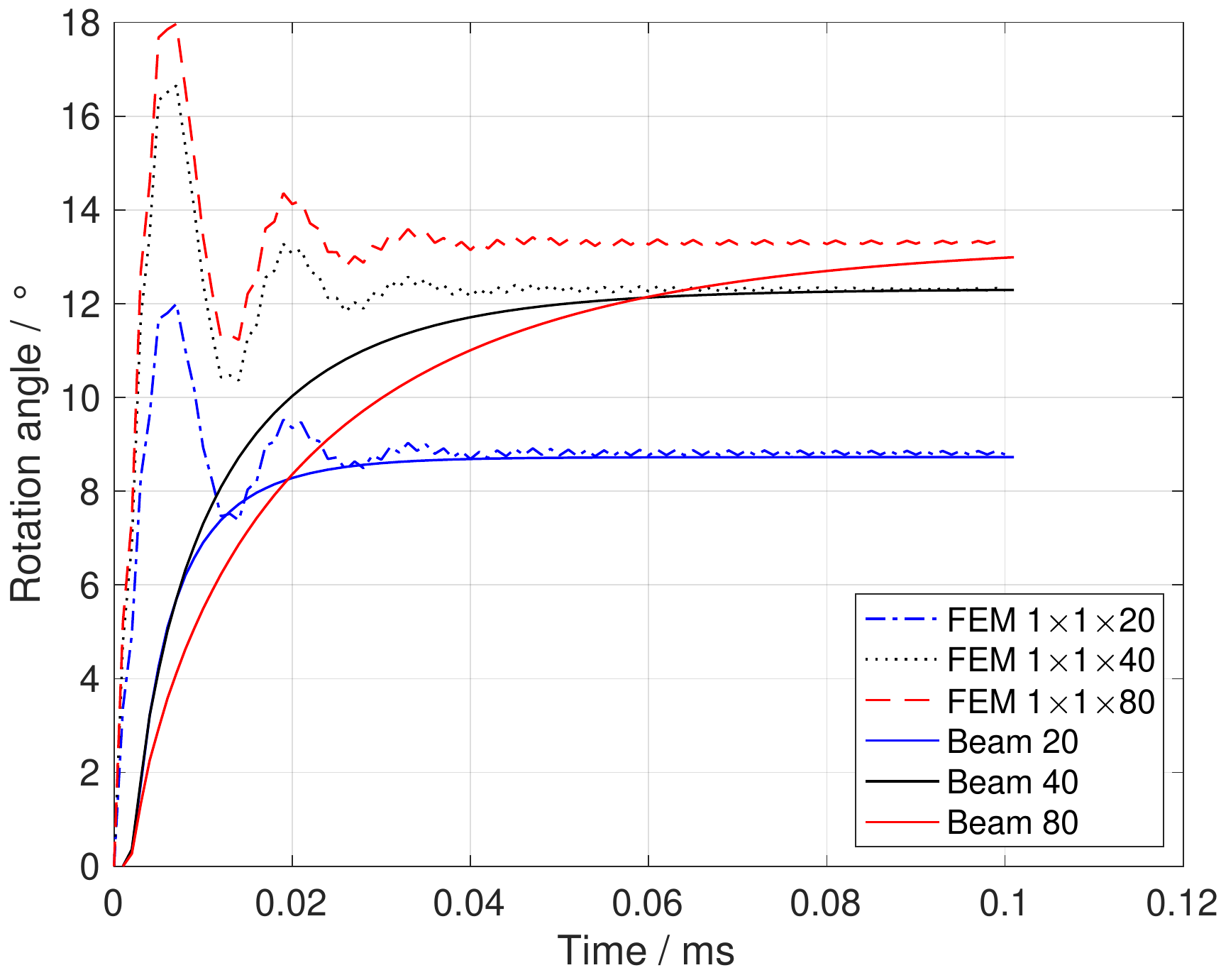}
	\caption{Rotation angle of the beam node at the free end.}
	\label{end_node3}
\end{figure}

The rotation angle of the beam node at the free end is shown in Fig. \ref{end_node3}, where the beam is discretized with 20, 40 and 80 elements in the {longitudinal} direction, respectively. {It can be observed that the rotation angles computed with the beam model and the 3D FEM model are very close at time $t=0.1ms$. With the finer mesh, the rotation angles are approaching to the converged value.} Additionally, the large deviation can be observed at the beginning of the loading, which is induced by the deformation of {the} cross section in the 3D FEM model during the rotation {and the approximation in deriving strain energy $\Omega_b$.}

\section{Conclusion}
In this paper, {an} electromechanically coupled viscoelastic beam model is developed. Based on the governing equations, the kinematics as well as the strain energy functions in continuum electromechanics, their counterparts in Cosserat beam are formulated consistently. Especially, the proposed formulation of {the} electric potential allows for all types of the dielectric induced deformations in the beam, such as contraction, shear, bending and torsion. In the uniaxial contractions of the beam, the oscillation of the beam is induced by the contractive electric forces, where the damping effect in the motion of the beam node is observed after introducing the viscoelastic effect. Additionally, the damping behavior of the total energy is observed as well, where the discrete Hamiltonian is evaluated. Another interesting point is that the oscillation frequency of the potential energy in the dielectric elastomer is twice as much as that of the displacement, which can be attributed to the fact that the potential energy is in its maximum value in the undeformed state of the charged dielectric elastomer. The simulation result also shows that the beam model agrees well with the 3D FEM model in contraction, shear, bending and torsion, however less degrees of freedom are required in {the} beam model. 
\section*{Acknowledgements}
The authors acknowledge the support of Deutsche Forschungsgemeinschaft (DFG) with the project: LE 1841-/5-11.

\section*{{Appendix}}
The aim of this appendix is to derive the strain energy function $\Omega_b$ for beam in Eq. (\ref{Ob}). The strain energy function $\Omega(\mathbf{C},J,\mathbf{E}^e)$ in Eq. (\ref{omega}) can be rewritten in terms of beam strain measures $\boldsymbol{\Gamma}, \mathbf{K},\boldsymbol{\gamma}, \boldsymbol{\kappa}, \boldsymbol{\varepsilon},\mathbf{e}$
\begin{align}
\Omega(\boldsymbol{\Gamma}, \mathbf{K},\boldsymbol{\gamma}, \boldsymbol{\kappa}, \boldsymbol{\varepsilon},\mathbf{e},\mathbf{X})=&\Omega(\mathbf{C},J,\mathbf{E}^e) \nonumber\\
=&\underbrace{\frac{\mu}{2} \left[ 2(\mathbf{a}^r \cdot \mathbf{d}^0_3)+ \mathbf{a}^r \cdot \mathbf{a}^r \right] - \mu  {\rm ln} (1+\mathbf{a}\cdot \mathbf{d}_3^t) + \frac{\lambda}{2} \left[ {\rm ln} (1+\mathbf{a}\cdot \mathbf{d}_3^t)\right] ^2}_{\Omega_1} \nonumber\\
&+ c_1 \left[  \varepsilon_1^2 + \varepsilon_2^2 + (\varepsilon_3+ \mathbf{e}\cdot\mathbf{X})^2  \right] + c_2 \left[  \varepsilon_1^2+ \varepsilon_2^2 + (1+\mathbf{a}^r \cdot \mathbf{a}^r) (\varepsilon_3 + \mathbf{e}\cdot\mathbf{X})^2 \right] + \nonumber\\
&\;\;\;\; \underbrace{c_2\left\lbrace 2 (\varepsilon_3 + \mathbf{e}\cdot\mathbf{X}) \left[  \varepsilon_1(\mathbf{d}_1^0 \cdot \mathbf{a}^r ) +\varepsilon_2(\mathbf{d}_2^0 \cdot \mathbf{a}^r ) +(\varepsilon_3 + \mathbf{e}\cdot\mathbf{X})(\mathbf{d}_3^0 \cdot \mathbf{a}^r)  \right]  \right\rbrace, \;\;\;\;}_{\Omega_2}
\end{align}
where $\mathbf{d}^0_i$ denotes $\mathbf{d}_i(s,0)$, $\mathbf{d}^t_i$ denotes $\mathbf{d}_i(s,t)$, $\varepsilon_i$ are the components of the strain-like vector $\boldsymbol\varepsilon$ in Eq. (\ref{estrain}), $\mathbf{X}$ is the vector $\mathbf{X}=\begin{bmatrix} X^1&X^2&0\end{bmatrix}$ and $\mathbf{e}$ is the electrical variable defined as
$\mathbf{e}(s)=\begin{bmatrix} \frac{\partial  \alpha^s}{\partial s} &   \frac{\partial  \beta^s}{\partial s} &  0\end{bmatrix}^T$.

The strain energy function for the beam is obtained by integrating $\Omega(\boldsymbol{\Gamma}, \mathbf{K},\boldsymbol{\gamma}, \boldsymbol{\kappa}, \boldsymbol{\varepsilon},\mathbf{e},\mathbf{X})$ over the cross section
\begin{align}
\Omega_b (\boldsymbol{\Gamma}, \mathbf{K},\boldsymbol{\gamma}, \boldsymbol{\kappa}, \boldsymbol{\varepsilon},\mathbf{e}) =\int_{\Sigma} \Omega(\boldsymbol{\Gamma}, \mathbf{K},\boldsymbol{\gamma}, \boldsymbol{\kappa}, \boldsymbol{\varepsilon},\mathbf{e},\mathbf{X}) dA=\Omega_{b1}+\Omega_{b2}
\end{align}
with
\begin{align}
\Omega_{b1}=&\int_{\Sigma} \Omega_1 dA\nonumber\\
=&\int_{\Sigma} \left\lbrace \frac{\mu}{2} \left[ 2(\mathbf{a}^r \cdot \mathbf{d}^0_3)+ \mathbf{a}^r \cdot \mathbf{a}^r\right] - \mu  {\rm ln} (1+\mathbf{a}\cdot \mathbf{d}_3^t) + \frac{\lambda}{2} \left[ {\rm ln} (1+\mathbf{a}\cdot \mathbf{d}_3^t)\right] ^2 \right\rbrace  dA\nonumber\\
\approx & \int_{\Sigma} \left\lbrace \frac{\mu}{2} \left[ 2(\mathbf{a}^r \cdot \mathbf{d}^0_3)+ \mathbf{a}^r \cdot \mathbf{a}^r )\right] - \mu \left[ \mathbf{a}\cdot \mathbf{d}_3^t-\frac{(\mathbf{a}\cdot \mathbf{d}_3^t)^2}{2}\right]  + \frac{\lambda}{2} \left[ \mathbf{a}\cdot \mathbf{d}_3^t-\frac{(\mathbf{a}\cdot \mathbf{d}_3^t)^2}{2}\right]^2 \right\rbrace  dA\nonumber\\
=& C_1 J_{1111}  + C_2 J_{1122} + C_3J_{11} + C_4J_{2222} +C_5J_{22}  +C_6
\end{align}
 and
\begin{align}
\Omega_{b2}&=\int_{\Sigma} \Omega_2 dA\nonumber\\
& =\int_{\Sigma} c_1 \left[  \varepsilon_1^2 + \varepsilon_2^2 + (\varepsilon_3+ \mathbf{e}\cdot\mathbf{X})^2  \right]  dA+c_2 \int_{\Sigma} \left[  \varepsilon_1^2+ \varepsilon_2^2 + (1+\mathbf{a}^r \cdot \mathbf{a}^r) (\varepsilon_3 + \mathbf{e}\cdot\mathbf{X})^2 \right]+\nonumber\\
&\;\;\;\;\;2 (\varepsilon_3 + \mathbf{e}\cdot\mathbf{X}) \left[  \varepsilon_1(\mathbf{d}_1^0 \cdot \mathbf{a}^r ) +\varepsilon_2(\mathbf{d}_2^0 \cdot \mathbf{a}^r ) +(\varepsilon_3 + \mathbf{e}\cdot\mathbf{X})(\mathbf{d}_3^0 \cdot \mathbf{a}^r)  \right] dA\nonumber\\
&= C_7 J_{1111}  + C_8 J_{1122} + C_9J_{11} + C_{10}J_{2222} +C_{11}J_{22}  +C_{12},
\end{align}
where the area moments are given by $J_{1111}=\int_{\Sigma} X_1^4dA$, $J_{2222}=\int_{\Sigma} X_2^4dA $, $J_{1122}=\int_{\Sigma} X_1^2X_2^2dA$, $J_{11}=\int_{\Sigma} X_1^2dA$, $J_{22}=\int_{\Sigma} X_2^2dA$, other area moments are zero due to the symmetry of the cross section in this work. 

The coefficients in $\Omega_{b1}$ and $\Omega_{b2}$ are given by
\begin{align}
C_1&= \frac{\lambda}{8} \left[ d_{31} (d_{21}  \kappa_3 + d_{31}  \kappa_2) + d_{32} (d_{22}  \kappa_3 + d_{32}  \kappa_2) + d_{33} (d_{23}  \kappa_3 + d_{33}  \kappa_2)\right] ^4,\\[10pt]
C_2&=\frac{3}{4} \lambda \left[ d_{31} (d_{11}  \kappa_3 - d_{31}  \kappa_1) + d_{32} (d_{12}  \kappa_3 - d_{32}  \kappa_1) + d_{33} (d_{13}  \kappa_3 - d_{33}  \kappa_1)\right] ^2\nonumber \\
&\qquad\cdot \left[ d_{31} (d_{21}  \kappa_3 + d_{31}  \kappa_2) + d_{32} (d_{22}  \kappa_3 + d_{32}  \kappa_2) + d_{33} (d_{23}  \kappa_3 + d_{33}  \kappa_2)\right] ^2,\\[10pt]
C_3&=\frac{\mu}{2} \left[ (d_{21}^0 K_3 + d_{31}^0 K_2)^2 + (d_{22}^0 K_3 + d_{32}^0 K_2)^2 + (d_{23}^0 K_3 + d_{33}^0 K_2)^2\right] \nonumber\\ 
& \quad +\frac{\mu}{2} \left[ d_{31} (d_{21}  \kappa_3 + d_{31}  \kappa_2) + d_{32} (d_{22}  \kappa_3 + d_{32}  \kappa_2) + d_{33} (d_{23}  \kappa_3 + d_{33}  \kappa_2)\right] ^2\nonumber\\
& \quad - \frac{\lambda}{2}  \left[ d_{31} (d_{21}  \kappa_3 + d_{31}  \kappa_2) + d_{32} (d_{22}  \kappa_3 + d_{32}  \kappa_2) + d_{33} (d_{23}  \kappa_3 + d_{33}  \kappa_2)\right] ^2\nonumber\\
&\quad \qquad \cdot \left[ d_{31} \gamma_1 + d_{32} \gamma_2 + d_{33} \gamma_3 - \frac{(d_{31} \gamma_1 + d_{32} \gamma_2 + d_{33} \gamma_3)^2}{2}\right]\nonumber\\
& \quad+ \frac{\lambda}{2} \left\lbrace d_{31} (d_{21}  \kappa_3 + d_{31}  \kappa_2)\right.\nonumber\\
&\quad \qquad  \left. - \left[ d_{31} (d_{21}  \kappa_3 + d_{31}  \kappa_2) + d_{32} (d_{22}  \kappa_3 + d_{32}  \kappa_2) + d_{33} (d_{23}  \kappa_3 + d_{33}  \kappa_2)\right]  (d_{31} \gamma_1 + d_{32} \gamma_2 + d_{33} \gamma_3) \right.\nonumber\\
&\quad \qquad \left.+ d_{32} (d_{22}  \kappa_3 + d_{32}  \kappa_2)+ d_{33} (d_{23}  \kappa_3 + d_{33}  \kappa_2)\right\rbrace^2,
\end{align}
\begin{align}
C_4&=\frac{\lambda}{8} \left[ d_{31} (d_{11}  \kappa_3 - d_{31}  \kappa_1) + d_{32} (d_{12}  \kappa_3 - d_{32}  \kappa_1) + d_{33} (d_{13}  \kappa_3 - d_{33}  \kappa_1)\right] ^4,\\[10pt]
C_5&=\frac{\mu}{2}\left[ (d^0_{11} K_3 - d^0_{31} K_1)^2 + (d^0_{12} K_3 - d^0_{32} K_1)^2 + (d^0_{13} K_3 - d^0_{33} K_1)^2\right]\nonumber\\
&\quad +\frac{\mu}{2} \left[ d_{31} (d_{11}  \kappa_3 - d_{31}  \kappa_1) + d_{32} (d_{12}  \kappa_3 - d_{32}  \kappa_1) + d_{33} (d_{13}  \kappa_3 - d_{33}  \kappa_1)\right] ^2\nonumber \\
&\quad - \frac{\lambda}{2} \left[ d_{31} (d_{11}  \kappa_3 - d_{31}  \kappa_1) + d_{32} (d_{12}  \kappa_3 - d_{32}  \kappa_1) + d_{33} (d_{13}  \kappa_3 - d_{33}  \kappa_1)\right]^2\nonumber\\
&\quad \qquad \cdot \left[ d_{31} \gamma_1 + d_{32} \gamma_2 + d_{33} \gamma_3 - \frac{(d_{31} \gamma_1 + d_{32} \gamma_2 + d_{33} \gamma_3)^2}{2}\right]\nonumber\\
 &\quad +  \frac{\lambda}{2} \left\lbrace  d_{31} (d_{11}  \kappa_3 - d_{31}  \kappa_1) - \left[ d_{31} (d_{11}  \kappa_3 - d_{31}  \kappa_1) + d_{32} (d_{12}  \kappa_3 - d_{32}  \kappa_1) + d_{33} (d_{13}  \kappa_3 - d_{33}  \kappa_1)\right] \right.\nonumber\\
&\quad \qquad \left. \cdot (d_{31} \gamma_1 + d_{32} \gamma_2 + d_{33} \gamma_3) + d_{32} (d_{12}  \kappa_3 - d_{32}  \kappa_1) + d_{33} (d_{13}  \kappa_3 - d_{33}  \kappa_1)\right\rbrace ^2, \\[10pt]
C_6&=\frac{\mu}{2}(\Gamma_1^2 + 2d^0_{31}\Gamma_1 + \Gamma_2^2 + 2d^0_{32}\Gamma_2 + \Gamma_3^2 + 2d^0_{33}\Gamma_3)\nonumber\\
&\quad + \mu \left[ d_{31} \gamma_1 + d_{32} \gamma_2 + d_{33} \gamma_3 - \frac{(d_{31} \gamma_1 + d_{32} \gamma_2 + d_{33} \gamma_3)^2}{2}\right]\nonumber\\
&\quad+\frac{\lambda}{2}\left[ d_{31} \gamma_1 + d_{32} \gamma_2 + d_{33} \gamma_3 - \frac{(d_{31} \gamma_1 + d_{32} \gamma_2 + d_{33} \gamma_3)^2}{2}\right] ^2,\\
C_7&=c_2 e_1^2 \left[ (d^0_{21} K_3 + d^0_{31} K_2)^2 + (d^0_{22} K_3 + d^0_{32} K_2)^2 + (d^0_{23} K_3 + d^0_{33} K_2)^2\right],
\end{align}
\begin{align}
C_8&=c_2 \left\lbrace e_1^2 \left[ (d^0_{11} K_3 - d^0_{31} K_1)^2 + (d^0_{12} K_3 - d^0_{32} K_1)^2 + (d^0_{13} K_3 - d^0_{33} K_1)^2\right]  \right.\nonumber\\
&\quad \qquad \left. + e_2^2 \left[ (d^0_{21} K_3 + d^0_{31} K_2)^2 + (d^0_{22} K_3 + d^0_{32} K_2)^2 + (d^0_{23} K_3 + d^0_{33} K_2)^2\right]  \right.\nonumber\\
&\quad \qquad \left. - 2 e_1 e_2 \left[ 2 (d^0_{11} K_3 - d^0_{31} K_1) (d^0_{21} K_3 + d^0_{31} K_2) + 2 (d^0_{12} K_3 - d^0_{32} K_1) (d^0_{22} K_3 + d^0_{32} K_2)\right]  \right.\nonumber\\
&\quad \qquad \left. -2e_1e_2 \left[ 2 (d^0_{13} K_3 - d^0_{33} K_1) (d^0_{23} K_3 + d^0_{33} K_2)\right] \right\rbrace ,\\[10pt]
C_9&=c_2\varepsilon_3^2 \left[ (d^0_{21} K_3 + d^0_{31} K_2)^2 + (d^0_{22} K_3 + d^0_{32} K_2)^2 + (d^0_{23} K_3 + d^0_{33} K_2)^2\right]\nonumber \\
&\quad +c_2 e_1^2 (\Gamma_1^2 + \Gamma_2^2 + \Gamma_3^2 + 1) \nonumber\\ 
& \quad + 2c_2 e_1 \varepsilon_1 \left[ d^0_{11} (d^0_{21} K_3 + d^0_{31} K_2) + d^0_{12} (d^0_{22} K_3 + d^0_{32} K_2) + d^0_{13} (d^0_{23} K_3 + d^0_{33} K_2)\right] \nonumber \\ 
& \quad+ 2 c_2e_1 \varepsilon_2 \left[ d^0_{21} (d^0_{21} K_3 + d^0_{31} K_2) + d^0_{22} (d^0_{22} K_3 + d^0_{32} K_2) + d^0_{23} (d^0_{23} K_3 + d^0_{33} K_2)\right]  \nonumber\\ 
& \quad + 2 c_2e_1 \varepsilon_3 \left[ d^0_{31} (d^0_{21} K_3 + d^0_{31} K_2) + d^0_{32} (d^0_{22} K_3 + d^0_{32} K_2) + d^0_{33} (d^0_{23} K_3 + d^0_{33} K_2)\right]\nonumber\\ 
& \quad + 2 c_2e_1 e_1 (d^0_{31} \Gamma_1 + d^0_{32} \Gamma_2 + d^0_{33} \Gamma_3) \nonumber\\
& \quad + 2c_2 \varepsilon_3 e_1 (d^0_{31} (d^0_{21} K_3 + d^0_{31} K_2) + d^0_{32} (d^0_{22} K_3 + d^0_{32} K_2) + d^0_{33} (d^0_{23} K_3 + d^0_{33} K_2)) \nonumber\\ 
& \quad + 2c_2 \varepsilon_3 e_1 (2 \Gamma_1 (d^0_{21} K_3 + d^0_{31} K_2) + 2 \Gamma_2 (d^0_{22} K_3 + d^0_{32} K_2) + 2 \Gamma_3 (d^0_{23} K_3 + d^0_{33} K_2)) \nonumber\\
& \quad + c_1 e_1^2,\\[10pt]
C_{10}&=c_2 e_2^2 \left[ (d^0_{11} K_3 - d^0_{31} K_1)^2 + (d^0_{12} K_3 - d^0_{32} K_1)^2 + (d^0_{13} K_3 - d^0_{33} K_1)^2\right],
\end{align}
\begin{align}
C_{11}&=c_1 e_2^2 \nonumber\\
&- 2c_2 e_2 \varepsilon_1 \left[ d^0_{11} (d^0_{11} K_3 - d^0_{31} K_1) + d^0_{12} (d^0_{12} K_3 - d^0_{32} K_1) + d^0_{13} (d^0_{13} K_3 - d^0_{33} K_1)\right]\nonumber \\
& -2c_2 e_2 \varepsilon_2 \left[ d^0_{21} (d^0_{11} K_3 - d^0_{31} K_1) + d^0_{22} (d^0_{12} K_3 - d^0_{32} K_1) + d^0_{23} (d^0_{13} K_3 - d^0_{33} K_1)\right] \nonumber\\
&- 2c_2 e_2 \varepsilon_3 \left[ d^0_{31} (d^0_{11} K_3 - d^0_{31} K_1) + d^0_{32} (d^0_{12} K_3 - d^0_{32} K_1) + d^0_{33} (d^0_{13} K_3 - d^0_{33} K_1)\right] \nonumber\\
& + 2c_2 e_2 e_2 (d^0_{31} \Gamma_1 + d^0_{32} \Gamma_2 + d^0_{33} \Gamma_3) \nonumber\\
& + c_2 e_2^2 (\Gamma_1^2 + \Gamma_2^2 + \Gamma_3^2 + 1)\nonumber\\
& + c_2 \varepsilon_3^2 \left[ (d^0_{11} K_3 - d^0_{31} K_1)^2 + (d^0_{12} K_3 - d^0_{32} K_1)^2 + (d^0_{13} K_3 - d^0_{33} K_1)^2\right] \nonumber \\
& -2c_2 \varepsilon_3 e_2 \left[ d^0_{31} (d^0_{11} K_3 - d^0_{31} K_1) + d^0_{32} (d^0_{12} K_3 - d^0_{32} K_1) + d^0_{33} (d^0_{13} K_3 - d^0_{33} K_1)\right] \nonumber \\
& -2c_2 \varepsilon_3 e_2 \left[ 2 \Gamma_1 (d^0_{11} K_3 - d^0_{31} K_1) + 2 \Gamma_2 (d^0_{12} K_3 - d^0_{32} K_1) + 2 \Gamma_3 (d^0_{13} K_3 - d^0_{33} K_1)\right], \\[10pt]
C_{12}&=c_1 (\varepsilon_1^2 + \varepsilon_2^2 + \varepsilon_3^2)\nonumber\\
& + 2c_2\varepsilon_3 \left[ \varepsilon_1 (d^0_{11} \Gamma_1 + d^0_{12} \Gamma_2 + d^0_{13} \Gamma_3) + \varepsilon_2 (d^0_{21} \Gamma_1 + d^0_{22} \Gamma_2 + d^0_{23} \Gamma_3) + \varepsilon_3 (d^0_{31} \Gamma_1 + d^0_{32} \Gamma_2 + d^0_{33} \Gamma_3)\right] \nonumber\\
& + c_2 \left[ \varepsilon_1^2 + \varepsilon_2^2 + \varepsilon_3^2 (\Gamma_1^2 + \Gamma_2^2 + \Gamma_3^2 + 1)\right].
\end{align}
In the above formulations, $(\cdot)_i(i=1,2,3)$ denote the components of the vector $(\cdot)$, such as $e_1$ is the first component of the vector $\mathbf{e}$ and $d_{12}^0$ is the second component of the vector $\mathbf{d}_1(s,0)$.

\clearpage
\small

\bibliographystyle{plainnat}
\bibliography{literature}

\end{document}